\documentclass[final,twocolumn,12p]{elsarticle}

\usepackage[sort&compress,comma,authoryear]{natbib}
\usepackage[colorlinks=true,urlcolor=blue,citecolor=red,linkcolor=red,bookmarks=true]{hyperref}
\setcitestyle{citesep={;}}

\usepackage{graphicx}
\usepackage{epstopdf}

\usepackage{booktabs}
\usepackage{multirow}
\usepackage[table,xcdraw]{xcolor}
\usepackage{color}
\usepackage{amsmath,amssymb,amsfonts}
\usepackage{algorithmic}
\usepackage{graphicx}
\usepackage{textcomp}
\usepackage{float}
\usepackage{xcolor}
\usepackage{pgf} 
\usepackage{pgfplots}
\usepackage{times}
\usepackage{adjustbox}

\usepackage[normalem]{ulem} 
\usepackage{pdfpages}
\usepackage[skins]{tcolorbox}
\usepackage{subfigure}
\usepackage[boxed]{algorithm2e}

\usepackage{tikz}
\usetikzlibrary{fit,calc}
\usepgfplotslibrary{external}

\SetKwProg{Fn}{Function}{}{end}\SetKwFunction{FRecurs}{FnRecursive}%

\def\BibTeX{{\rm B\kern-.05em{\sc i\kern-.025em b}\kern-.08em
    T\kern-.1667em\lower.7ex\hbox{E}\kern-.125emX}}

\usepackage{array}
\newcolumntype{C}[1]{>{\centering\arraybackslash}p{#1}}

\usetikzlibrary{arrows,automata}

\journal{Expert Systems with Applications}

\begin{document}

\newenvironment{RQ}{\vspace{2mm}\begin{tcolorbox}[enhanced,width=3.0in,size=fbox,fontupper=\normalsize,colback=blue!5,drop shadow southwest,sharp corners]}{\end{tcolorbox}}



\title{Understanding Static Code Warnings: an Incremental AI Approach}

\author{Xueqi Yang}
\ead{xyang37@ncsu.edu}
\author{Zhe Yu}
\ead{zyu9@ncsu.edu}
\author{Junjie Wang}
\ead{wangjunjie@itechs.iscas.ac.cn}
\author{Tim Menzies}
\ead{tim.menzies@gmail.com}

\address{Department of Computer Science, North Carolina State University, Raleigh, NC, USA}
\address{Institute of Software Chinese Academy of Sciences, Beijing, China}

\begin{abstract}

Knowledge-based systems reason over some knowledge base. 
Hence, an important issue for such systems is how to acquire the knowledge needed
for their inference.
This paper assesses active learning methods for acquiring knowledge for ``static code warnings''.

Static code analysis is a widely-used method
for detecting bugs and security vulnerabilities in software systems. As software becomes more complex, analysis tools also report lists of increasingly complex warnings that developers need to address on a daily basis.  
Such static code analysis tools
are usually over-cautious; i.e. they often offer many warnings about spurious issues.
Previous research work shows that about 35\% to 91 \% warnings reported as bugs by SA tools are actually unactionable~ (i.e., warnings that would not be acted on by developers because they are falsely suggested as bugs).

Experienced developers know which errors are important and which can be safely ignored. How can we capture that experience?
This paper reports on an incremental AI tool that watches humans reading false alarm reports. Using an incremental support vector machine mechanism, this AI tool can quickly learn to distinguish spurious false alarms from more serious matters that deserve further attention.

 In this work, nine open-source projects are employed to evaluate our proposed model on the features extracted by previous researchers and identify the actionable warnings in a priority order given by our algorithm. We observe that our model can identify over 90\% of actionable warnings when our methods tell humans to ignore 70 to 80\% of the warnings.
 

\end{abstract}

\begin{keyword}
Actionable warning identification\sep Active learning\sep Static analysis\sep Selection process
\end{keyword}
\maketitle

\section{Introduction}
\label{sec:intro}






Knowledge acquisition problem is a longstanding and challenging bottleneck in artificial intelligence, especially like Semantic Web project~\citep{feigenbaum1980knowledge}. Traditional knowledge engineering methodologies
handcraft the knowledge prior to testing that data
on some domain~\citep{hoekstra2010knowledge}.
Such handcrafted knowledge is expensive to collect. Also, building competent
systems can require extensive manually crafting-- which leads to a long gap between crafting and testing knowledge.

\begin{figure*}[!t]
\begin{center}
\includegraphics[width=4.5in]{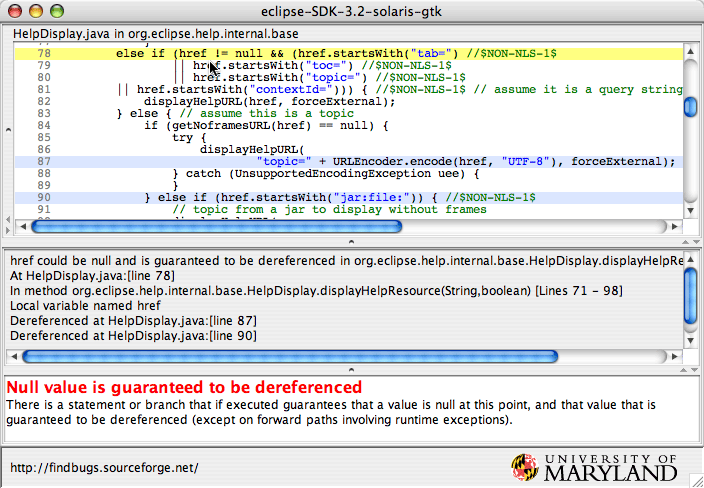}\end{center}
\caption{Example of a static code analysis warning, generated via the FindBugs
tool.}\label{fig:fb}
\end{figure*}

In this paper, we address these problems with self-adaptive incremental active learning utilizing a human-in-the-loop process.
This approach can be leveraged to filter spurious vs serious static warnings generated by static analysis (SA) tools. 

Static code analysis is a common operation of detecting bugs and security vulnerabilities in software systems. 
The wide range of commercial applications of static analysis demonstrates the industrial perception that these tools have a very high economic value.
One of the popular SA tools, FindBugs\footnote{http://findbugs.sourceforge.net/} (shown in Figure~\ref{fig:fb}) has been downloaded over a million times so far. 
However, large amounts of warnings are falsely suggested by SA tools as bugs to developers, overwhelming the few actionable ones (true bugs). Due to high rates of unactionable warnings, the utility of such static code analysis tools is questionable.
Previous research work shows that about 35\% to 91 \% warnings reported as bugs by SA tools are actually unactionable~ \citep{kim2007warnings,heckman2008establishing,heckman2011systematic}.

Experienced developers have the knowledge of filtering out ignorable and unactioable warnings.
Our active learning methods incrementally acquire and validate this knowledge.
By continuously and incrementally constructing and updating the model, our approach can help SE developers to identify more actionable static warnings with very low inspection costs and provide an efficient way to deal with software mining on the early life cycle.  

This paper evaluates the proposed approach with the following four research questions:

\begin{RQ}
{\bf RQ1.} What is the baseline rate for bad static warnings?
\end{RQ}

While this is more a systems question rather than a research question, it is a necessary precondition to our work since it documents the problem we are trying to address. For this question, we report results from FindBugs. These results will serve as the baseline for the rest of our work.

\begin{RQ}

{\bf RQ2.} What is the previous state-of-the-art method to tackle the prevalence of actionable warnings in SA tools?
\end{RQ}

Wang et al.~\citep{wang2018there} conduct a systematic evaluation of all the publicly available features (116 features in total) that discuss static code warnings.
That work offered a "golden set of features"; i.e., 23 features that Wang et al.~\citep{wang2018there} argued were most useful for extracting serious bug reports generated from FindBugs.
Our experiments combining three supervised learning models from the literature with these 23 features.


\begin{RQ}
{\bf RQ3.} Does incremental active learning reduce the cost to identify actionable Static Warnings?
\end{RQ}

We will show that incremental active learning reduces the cost of identifying actionable warnings dramatically (and obtains performance almost as good as supervised learning).

\begin{RQ}
{\bf RQ4.} How many samples should be retrieved to identify all the actionable Static Warnings?
\end{RQ}

In this case study, incremental active learning can identify over 90\% of actionable warnings by learning from about 20\% to 30\% of data.  
Hence, we recommend this system to developers who wish to reduce the time they waste chasing spurious errors.



\subsection{Organization of this Paper}
The remainder of this paper is organized as follows. Research background and related work is introduced in Section~\ref{sec:related}. In Section~\ref{sec:method}, we describe the detail of our methodology. Our experiment details are introduced in Section~\ref{sec:experiment}. In Section~\ref{sec:evaluation}, we answer proposed research questions. Threats to validity and future work are discussed in Section~\ref{sec:threats} and we finally draw a conclusion in Section~\ref{sec:conclusion}.

To facilitate other researchers in this area, all our scripts are data are freely available on-line\footnote{Download our scripts and data from \url{https://github.com/XueqiYang/incrementally-active-learning_SWID}.}.

\subsection{Contributions of this Paper}

In the literature, 
active learning methods have been extensively discussed, like finding relevant papers in literature review~\citep{yu2018finding,yu2019fast2}, security vulnerability prediction~\citep{8883076}, crowdsourced testing~\citep{wang2016local}, place-aware application development~\citep{murukannaiah2015platys}, classification of software behavior~\citep{bowring2004active}, and multi-objective optimization~\citep{krall2015gale}. The unique contribution of this work lies in the novel application of these methods to resolving problems with static code warnings. To the best of our knowledge, no prior work has tried to tame spurious static code warnings by treating these as an incremental knowledge acquisition problem.

\section{Related Work}
\label{sec:related}
\subsection{Reasoning About Source Code}

The software development community has produced numerous static code analysis tools such as FindBugs, PMD\footnote{https://pmd.github.io/}, or Checkstyle\footnote{http://checkstyle.sourceforge.net/} that are able to generate various warnings to help developers identifying potential code problems. Such static code analysis tools such as FindBugs leverage \textit{static analysis} (SA) techniques to inspect source code for the occurrence of bug patterns (i.e., the code idiom that is often an error) without actually executing nor considering an exact input. These bugs detected by FindBugs are grouped into a pattern list, (i.e, performance, style, correctness and so forth) and each bug is reported by FindBugs with priority from 1 to 20 to measure the severity, which is finally grouped into four scales either scariest, scary, troubling, and of concern~\citep{ayewah2008using}.



Some SA tools learn to identify new bugs using historical data from past problems.
This is not ideal since it means that whenever there are chances to tasks, languages, platforms, and perhaps even developers then the old warnings might go out of date and new ones have to be learned.
Static warning identification is increasingly relying on complex software systems~\citep{wijayasekara2012mining}. Identifying static warnings in every stage of the software life cycle is essential, especially for projects in early development stage~\citep{murtaza2016mining}.

Arnold et al.~\citep{arnold2009security} suggests that every project, early in its own lifecycle, should build its own static warning system.
Such advice is hard to follow since it means a tedious, time-consuming and expensive retraining process at the start of each new project. To say that in another way, Arnold et al.'s advice suffers from the knowledge acquisition bottleneck problem.

\subsection{Static Warning Identification}

Static warning identification aims at identifying common coding problems early in the development process via SA tools and distinguishing actionable warnings from unactionable ones~\citep{heckman2011systematic,hovemeyer2004finding,yan2017revisiting}. 


Previous studies have shown that false positives in static alerts have been one of the most important barriers for developers to use static analysis tools~\citep{thung2015extent,avgustinov2015tracking,johnson2013don}. To address this issue, many techniques have been introduced to identify actionable warnings or alerts.
Various models have been mentioned in their study, including graph theory~\citep{boogerd2008assessing,bhattacharya2012graph}, machine learning~\citep{wang2016automatically,shivaji2009reducing} etc. However, most of the studies are plagued by a common issue, choosing the appropriate warning characteristics from abundant feature artifacts proposed by SA studies so far. 


Ranking schemes are one way to improve
static analysis tool~\citep{kremenek2004correlation}.  Allier et al.~\citep{allier2012framework} proposed a framework to compare 6 warning ranking algorithms and identified the best one to rank warnings. Similarly, Shen et al.~\citep{shen2011efindbugs} employed a ranking technique to rank the true error reports on top so as to reduce false positive warnings. Some other works also prioritize warnings by selecting different categories of impact factors~\citep{liang2010automatic} or by analyzing software history~\citep{kim2007prioritizing}.

Recent work has shown that this problem can be solved by combining machine learning techniques to identify whether a detected warning is actionable or not, e.g., finding alerts with similar code patterns and building prediction models to classify new alerts~\citep{hanam2014finding}. Heckman and Williams did a systematic literature review revealing that most of these works focus on exploring a reasonable characteristic set, like Alert characteristics (AC) and Code characteristics (CC), to distinguish actionable and unactionable warnings more accurately~\citep{heckman2011systematic,hanam2014finding,heckman2009model}. One of the most integrated study explores 15 machine learning algorithms and 51 warning characteristics derived from static analysis tools and achieves good performance with high recall (83-99 \%)~\citep{heckman2009model}. However, in practice, information on bug warning patterns is limited to be obtained, especially for some trivial checkers in SA tools. Also, these tools suffer from conflation issues where similar warnings are given different names in different studies.



Wang et al.~\citep{wang2018there} recently conducted a systematic literature review to collect all publicly available features (116 in total) for SA analysis and implemented a tool based on Java for feature extraction. All the values of these collected features are extracted from warning reports generated by FindBugs based on 60 revisions of 12 projects. Six machine learning classifiers were employed to automatically identify actionable static warning. 23 common features were identified as the best and most useful feature combination for Static Warning Identification, since the best performance is always obtained when using these 23 golden features, better than using total feature set or other subset strategies. To the best of our knowledge, this is the most exhaustive research about SA characteristics yet published.






\subsection{Active Learning}

Labeled data is required by supervised machine learning techniques. Without such data, these algorithms cannot learn predictors.
Obtaining good labeled data can sometimes be time consuming and expensive. In the case of this paper, we are concerned with learning how to label static code warnings (spurious or serious). For another example, training a good document classifier might require hundreds of thousands of samples. Usually, these examples do not come with labels, and therefore expert knowledge (e.g., recognizing a handwritten digit) is required to determine the ``right'' label.  

Active learning~\citep{settles2009active} is a machine learning algorithm that enables the learners to actively choose which examples to label from amongst the currently unlabeled instances. This approach trains on a little bit of labeled data, and then asks again for some more labels for the unlabelled examples that are most ``interesting'' (e.g. whose labels are most uncertain). This process greatly reduces the amount of labeled data required to train a model while still achieving good predictive performance.

Active learning has been applied successfully in several SE research areas, such as finding relevant papers in literature review~\citep{yu2018finding,yu2019fast2}, security vulnerability prediction~\citep{8883076}, crowd sourced testing~\citep{wang2016local}, place-aware application development~\citep{murukannaiah2015platys}, classification of software behavior~\citep{bowring2004active}, and multi-objective optimization~\citep{krall2015gale}.
Overall, there are three different categories of active learning:

\begin{itemize}
    \item \textit{Membership query synthesis.} In this scenario, a learner is able to generate synthetic data for labeling, which might not be applicable to all cases. 
    \item \textit{Stream-based selective sampling.} Each sample is considered separately in the case of label querying or rejection. There are no assumptions on data distribution, and therefore it is adaptive to change.
    \item \textit{Pool-based sampling.} Samples are chosen from a pool of unlabeled data for the purpose of labeling. The learner is usually initially trained on a fully labeled fraction of data to generate a preliminary model, which is subsequently used to identify which sample would be most beneficial to be used next in the training set during the next generation of active learning loop. Pool-based sampling scenario is the most widely adopted scheme in literature, which is also applied in our work.
\end{itemize}

Previous work has shown the successful adoption of active learning in several research areas. 
Wang et.al~\citep{wang2016local} applied active learning to identify the test reports that reveal ``true fault'' from a large amount of test reports in crowdsourced testing of GUI applications. 
Within that framework, they proposed a classification technique that labels a fraction of most informative samples with user knowledge, and trained classifiers based on the local neighborhood.

Yu et.al~\citep{yu2018finding,yu2019fast2,yu2019searching} proposed a framework called FASTREAD to assist researchers to find the relevant papers to read. FASTREAD works by 1) leveraging external domain knowledge (e.g., keyword search) to guide the initial selection of papers; 2) using an estimator of the number of remaining paper to decide when to stop; 3) applying error correction algorithm to correct human mislabeling. This framework has also been shown effective in solving other software engineering problems~\citep{yu2018total}, such as inspecting software security vulnerabilities~\citep{8883076}, finding self-admitted technical debt~\citep{fahid2019better}, and test case prioritization~\citep{yu2019terminator}. In this work, we adopt a similar framework in static warning analysis.

To the best of our knowledge, this work is the first study to utilize incremental active learning to reduce unnecessary inspection of static warnings based on the most effective feature attributes.
While Wang et al. is the closest work to this paper, we differ very much from their work.
\begin{itemize}
\item
In that study, their raw data was screen-snaps of erroneous conditions within a GUI. 
Also, they spend much effort tuning a feedback mechanism specialized for their images.
\item
In our work, our raw data is all textual (the  text of a static code warning).
We found that a different method, based on active learning, worked best for such textual data.
\end{itemize}

\section{Methodology}
\label{sec:method}

\subsection{Overview}

\begin{figure}[!b]
\center{\includegraphics[width=0.45\textwidth]{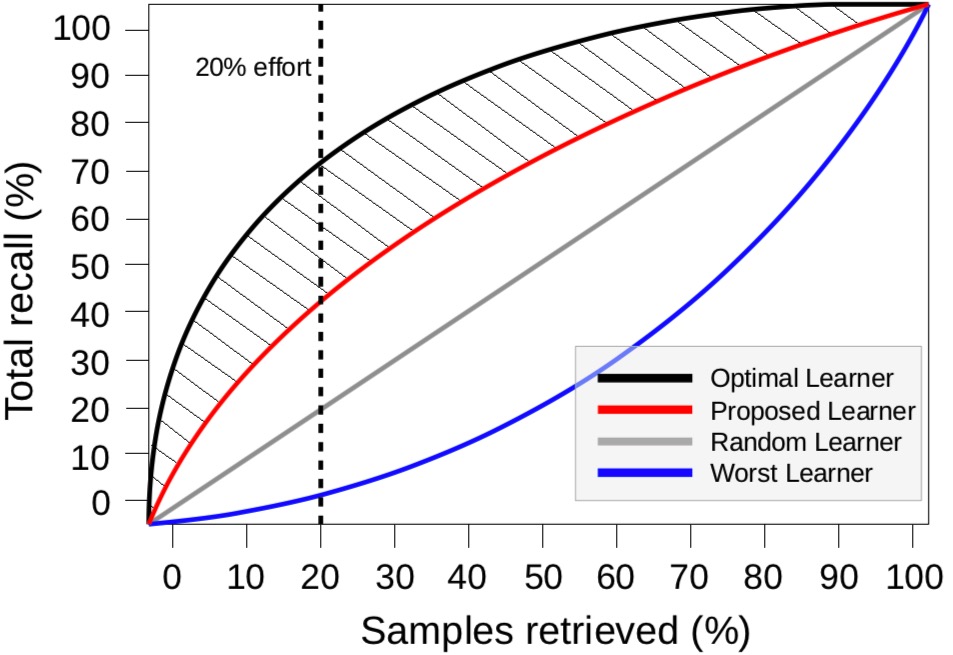}}
\caption{Learning Curve of Different Learners. }
\label{fig:learning}
\end{figure}

This work applies an incremental active learning framework to identify static warnings. 
This is derived from active learning, which has been proved outperformed in solving the total recall problem in several areas, e.g., electronic discovery, evidence-based medicine, primary study selection, test case prioritization, and so forth.
As illustrated in Figure ~\ref{fig:learning}, we aim to achieve higher recall with lower effort in inspecting warnings generated by SA tools.


\subsection{Evaluation Metrics}
\label{subsec:metrics}

Table \ref{table:despription} represents all the variables involved in our study.
We evaluated the active learning results in terms of \textbf{\textit{total recall}} and \textbf{\textit{cost}}, which are demonstrated as follows:



\begin{table}[!htbp]
\small
\caption{Description of Variables in Incremental Active Learning.}
\begin{adjustbox}{max width=0.48\textwidth}
\begin{tabular}{l|l}
\hline
\multicolumn{1}{c|}{\textbf{Variable}} & \multicolumn{1}{c}{\textbf{Description}} \\ \hline
$E$ & \begin{tabular}[c]{@{}l@{}}Set of warning that reported by static \\ analysis tools\end{tabular} \\ \hline
$T$ & \begin{tabular}[c]{@{}l@{}}Set of actionable warning or target\\ samples\end{tabular} \\ \hline
$L$ & \begin{tabular}[c]{@{}l@{}}Set of warning that has been currently\\ retrieved or labeled\end{tabular} \\ \hline
$L_T$ & \begin{tabular}[c]{@{}l@{}}Set of warning has been currently labeled\\ and reveals actionable warning\end{tabular} \\ \hline
\textbf{Total Recall} & $L_T / T$ \\ \hline
\textbf{cost} & $ L $/$ E$ \\ \hline
\end{tabular}
\label{table:despription}
\end{adjustbox}
\end{table}

\textit{Total recall} addresses the ratio between samples labeled but not revealing actionable warning and total real actionable warning samples.
The optimal value of \textit{total recall} is 1, which represents all of the target samples (or actionable warning in our case) have been retrieved and labeled as actionable. 

\textit{Cost} considers the set of warning that has currently been retrieved or labeled out of the set of warning reported by the static warning analysis tools. The value of cost varies between the ratio of actionable warnings in the dataset and 1. The lower bound means active learning algorithm prioritizes all targeted samples without uselessly labeling any unactionable warnings. This is a theoretical optimal value (which, in practice,
may be unreachable).
The upper bound means active learning algorithm successfully retrieves all the real warning samples, but at the cost of labeling them all
(which is meaningless because randomly labeling samples will achieve the same goal).

Figure~\ref{fig:learning} is an Alberg diagram showing the learning curve of different learners. In this figure, the x-axis and y-axis respectively represent the percentage of warnings retrieved or labeled by learners (i.e. cost) and the percentage of actionable warnings retrieved out of total actionable ones (i.e. total recall). An optimal learner will achieve higher total recall than others when a specific cost threshold is given, e.g., at the cost of 20 \% effort as illustrated in Figure~\ref{fig:learning}. The best performance in Figure~\ref{fig:learning} is obtained by optimal learner, followed by proposed learner, random learner and worst learner. This learning curve is a performance measurement at different cost thresholds settings.  

\textit{AUC} (Area under the ROC Curve) measures the area under the Receiver Operator Characteristic (ROC)   curve~\citep{witten2016data, heckman2011systematic} and reflects the percentage of actionable warnings against the percentage of unactionable ones so as to overall report the discrimination of a classifier~\citep{wang2018there}. This is a widely adopted measurement in Software Engineering, especially for imbalanced data~\citep{liang2010automatic}.

\subsection{Active Learning Model Operators}
Several operators are apply to address the challenge of the total recall problem, as listed in Table~\ref{tbl:FASTREADOperator}. Specific details about each operator are illustrated as follows:

\begin{table}[!t]
\small
\centering
\caption{Operators of Active Learning.}
\begin{adjustbox}{max width=0.48\textwidth}
\begin{tabular}{l|l}
\hline
\multicolumn{1}{c|}{\textbf{Operator}} & \multicolumn{1}{c}{\textbf{Description}} \\ \hline
Machine Learning Classifier & \begin{tabular}[c]{@{}l@{}}Widely-used classification \\ technique.\end{tabular} \\ \hline
\begin{tabular}[c]{@{}l@{}}Presumptive non-relevant \\ examples\end{tabular} & \begin{tabular}[c]{@{}l@{}}Alleviate the sampling bias of \\ non-relevant examples.\end{tabular} \\ \hline
Aggressive Undersampling & Data-balancing technique. \\ \hline
Query strategy & \begin{tabular}[c]{@{}l@{}}Uncertainly sampling and \\ certainty sampling in active \\ learning.\end{tabular} \\ \hline
\end{tabular}
\label{tbl:FASTREADOperator}
\end{adjustbox}
\end{table}

\textbf{Classifier}
We employ three machine learning classifiers as an embedded active learning model, linear SVM with weighting scheme, Random Forest and Decision Tree with default parameters as these classifiers are widely explored in software engineering area and also reported in Wang et al.'s paper. All of the classifiers are modules from Sckit-learn~\citep{pedregosa2011scikit}, a Python package for machine learning.

\textbf{Presumptive non-relevant examples}, proposed by Cormack et al.~\citep{cormack2015autonomy}, is a technique to alleviate the sample bias of negative samples in unbalanced dataset. To be specific, before each training process, the model samples randomly from the unlabeled pool and assumes that the sampled instance is labeled as negative in training, due to the prevalence of negative samples.

\textbf{Aggressive undersampling} ~\citep{wallace2009meta} is a sampling method to cope with an unbalanced dataset by throwing away majority negative training points close to the decision plane of SVM and aggressively accessing minority positive points until the ratio of these two categories is balanced. It's an effective approach to kill unbalanced bias in datasets.
This technique is suggested by Wallace et al.~\citep{wallace2010semi} after the initial stage of incremental active learning and when the established model becomes stable.

The \textbf{querying strategy}
is the approach utilized to determine which data instance in an unlabelled pool to query for labelling next. We adopt two of the most commonly used strategies, \textit{uncertainty sampling} ~\citep{settles2009active} and \textit{certainty sampling}~\citep{miwa2014reducing}.

Uncertainty sampling~\citep{settles2009active}
is the simplest and most commonly used query strategy in active learning, where unlabeled samples closest to the decision plane of SVM or predicted to be the least likely positive by a classifier are sampled for query.  Wallace et al.~\citep{wallace2010semi} recommended uncertainty sampling method in biomedical literature review and it reduces the cost of manually screening literature efficiently.

Certainty sampling ~\citep{miwa2014reducing} is a kind of greedy algorithm to maximize the utility of incremental learning model by prioritizing the samples which are the most likely to be actionable warnings. Contrary to uncertainty sampling, certainty sampling method gives priority to the instances which are far away from the decision plane of SVM or have the highest probability score predicted by the classifier. It speeds up the process of retrieving and plays the major role of stopping earlier.

\begin{figure}[!b]
\centerline{\includegraphics[width=0.5\textwidth]{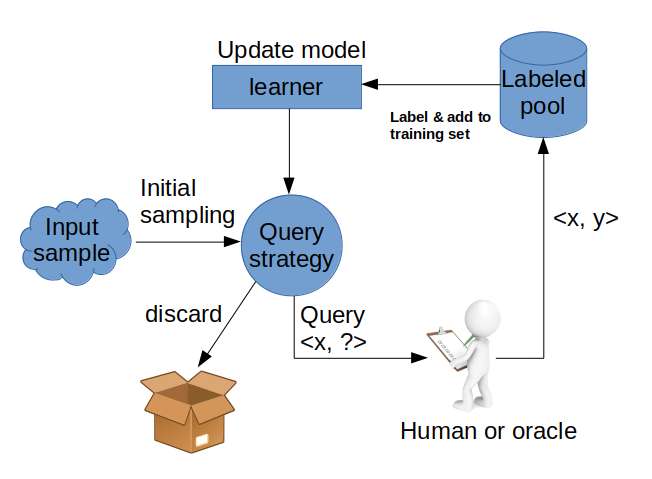}}
\caption{Procedure of Incremental Active Learning.}   
\label{fig:procedure}
\end{figure}

\subsection{Active Learning Procedures}

Figure ~\ref{fig:procedure} presents the procedures of incremental active learning, and a detailed description of each step is demonstrated as follows:



\begin{enumerate}
    \item Initial Sampling.
    
    We propose two initial sampling strategies to cope with the scenario that historical information is available or not.
    
    For software projects in early life cycle without sufficient historical revisions in the version control system, random sampling without replacement is used in the initial stage when the labeled warning pool is NULL.
    
    For software projects with previous version information, we utilize version N-1 to get a pre-trained model and initialize sampling on version N. This practice can reduce the cost of manually excluding unactionable warnings since the prevalence of false positive in SA datasets.
    
    
    \item Human or oracle labeling.
    
    After a warning message is selected by initial sampling or query strategy, manual inspection is required to identify whether the retrieved warning is actionable or not. In our simulation, the ground truth serves as a human oracle and it returns a label once a warning presupposed as unlabeled is queried by active learning model.
    
    In static analysis, inspecting and tagging the warning being queried is considered as a main overhead of this process. As demonstrated in Table \ref{tabel:variables}, this overhead is  denoted as \textbf{Cost} and is what software developers strive to reduce.
    
    \item Model Training and updating.
  
    After a new coming-in  warning is labeled by human oracle, this data sample is added to training data. The model is retrained and updated recursively.
    
    \item Query Strategy.
    
    Uncertainty sampling is leveraged when the actionable  samples retrieved and labeled by our model is under a specific threshold. This query strategy mainly applies when targeted data samples are rare in training set and building a stable model faster is required~\citep{8883076}.
    
    Finally, after labeled actionable warning  exceed the given threshold, certainty sampling is employed to aggressively searching for true positive and greedily reduce the cost of warning inspection.  
\end{enumerate}

\begin{table}[!htbp]
\footnotesize
\centering
\caption{Summary of Projects Surveyed.}
\tabcolsep=0.11cm
\begin{tabular}{l|l|l|l}
\hline
\multicolumn{1}{c|}{\textbf{Project}} & \multicolumn{1}{c|}{\textbf{Period}} & \multicolumn{1}{c|}{\textbf{\begin{tabular}[c]{@{}c@{}}Revision-\\ Interval\end{tabular}}} & \multicolumn{1}{c}{\textbf{Domain}} \\ \hline
\begin{tabular}[c]{@{}l@{}}Lucence\\-solr\end{tabular} & \begin{tabular}[c]{@{}l@{}}01/2013-\\ 01/2014\end{tabular} & 3 month & Search engine \\ \hline
Tomcat & \begin{tabular}[c]{@{}l@{}}01/2013-\\ 01/2014\end{tabular} & 3 month & Server \\ \hline
Derby & \begin{tabular}[c]{@{}l@{}}01/2013-\\ 01/2014\end{tabular} & 3 month & Database \\ \hline
Phoenix & \begin{tabular}[c]{@{}l@{}}01/2013-\\ 01/2014\end{tabular} & 3 month & Driver \\ \hline
Cassandra & \begin{tabular}[c]{@{}l@{}}01/2013-\\ 01/2014\end{tabular} & 3 month & Big data manage \\ \hline
Jmeter & \begin{tabular}[c]{@{}l@{}}01/2012-\\ 01/2014\end{tabular} & 6 month & Performance manage \\ \hline
Ant & \begin{tabular}[c]{@{}l@{}}01/2012-\\ 01/2014\end{tabular} & 6 month & Build manage \\ \hline
\begin{tabular}[c]{@{}l@{}}Commons\\ .lang\end{tabular} & \begin{tabular}[c]{@{}l@{}}01/2012-\\ 01/2014\end{tabular} & 6 month & Java utility \\ \hline
Maven & \begin{tabular}[c]{@{}l@{}}01/2012-\\ 01/2014\end{tabular} & 6 month & Project manage \\ \hline
\end{tabular}
\label{table:project}
\end{table}

\section{Experiment}
\label{sec:experiment}

\subsection{Static Warning Dataset}

The nine datasets explored in this work are collected from previous research.
Wang et al.~\citep{wang2018there} performed a systematic literature review to gather all publicly available features (116 in total) for SA analysis. 
For this research, all the values of this collected feature set were extracted from warnings reported by FindBugs on the 60 successive revisions of 12 projects. Using 
the Static Warning (SA) tool, we applied  FindBugs to 60 revisions from 12 projects' revision history. By collected performance statistics from three supervised learning classifiers on 12 datasets, a golden feature set (23 features) is found via a greedy backward elimination algorithm. We utilize the best feature combination as the warning characteristics in our research.

On closer inspection of these datasets, we found three projects with obvious data inconsistency issues (such as data features mismatch with data labels).
Hence, our study only explored the remaining nine projects.

Table \ref{table:project} lists the summary of projects surveyed in our paper. For each project, there are 5 versions collected from starting revision time after a specific revision interval. We train the model on version 4 and test on version 5.
 
Previous research~\citep{wang2018there} collected 116 static warning features with a systematic literature review. These features fall into eight categories, and 95 features are left after eliminating unavailable ones as shown in Table \ref{tabel:variables}. We employ the 23 golden features as the independent variables proposed by Wang et al., which are highlighted in bold in Table \ref{tabel:variables}.
In our study, the dependent variable is actionable or unactionable.
These labels were generated via a method proposed by previous researches~\citep{heckman2008establishing,hanam2014finding,liang2010automatic}. 
That is, for a specific warning, if it is closed in later revision after a revision interval when the project was collected, it will finally be labeled as actionable. For warning still existing after later revision interval, it will be labeled as unactionable. Otherwise, for some minority warnings which are deleted after later interval, they will be removed and ignored in our study.  

Table \ref{table:numofSamples} shows the number of warnings and distribution of each warning type (as reported by FindBugs) in nine software projects. Note that our data is highly imbalanced with the ratio of targeted samples from 3 to 34 percent. 

\begin{table}[]
\small
\caption{Categories of Selected Features.~(8 categories are shown in the left column, and 95 features explored in Wang et al. are shown in the right column with 23 golden features in bold.)}
\tabcolsep=0.11cm
\begin{adjustbox}{max width=0.48\textwidth}
\begin{tabular}{ll}
\hline
\textbf{Category}  &   \textbf{Features} \\ \hline
Warning combination & \begin{tabular}[c]{@{}l@{}}size content for warning type;\\ size context in method, file, package;\\ \textbf{warning context in method, file,} package;\\ \textbf{warning context for warning type};\\ fix, non-fix change removal rate;\\ \textbf{defect likelihood for warning pattern};\\ variance of likelihood;\\ defect likelihood for warning type;\\ \textbf{discretization of defect likelihood}; \\ \textbf{average lifetime for warning type};\end{tabular} \\ \hline
Code characteristics & \begin{tabular}[c]{@{}l@{}}method, file, package size;\\ comment length;\\ \textbf{comment-code ratio};\\ \textbf{method, file depth};\\ method callers, callees;\\ \textbf{methods in file}, package;\\ classes in file, \textbf{package};\\ indentation;\\ complexity;\end{tabular} \\ \hline
Warning characteristics & \begin{tabular}[c]{@{}l@{}}\textbf{warning pattern, type, priority,} rank;\\ warnings in method, file, \textbf{package};\end{tabular} \\ \hline
File history & \begin{tabular}[c]{@{}l@{}}latest file, package modification;\\ file, package staleness;\\ \textbf{file age}; \textbf{file creation};\\ deletion revision; \textbf{developers};\end{tabular} \\ \hline
Code analysis & \begin{tabular}[c]{@{}l@{}}call name, class, \textbf{parameter signature},\\ return type; \\ new type, new concrete type;\\operator;\\ field access class, field; \\catch;\\ field name, type, visibility, is static/final;\\ \textbf{method visibility}, return type,\\ is static/ final/ abstract/ protected;\\ class visibility, \\ is abstract / interfact / array class;\end{tabular} \\ \hline
Code history & \begin{tabular}[c]{@{}l@{}}added, changed, deleted, growth, total, percentage \\ of LOC in file in the past 3 months;\\ \textbf{added}, changed, deleted, growth, total, percentage \\ of LOC in file in the last 25 revisions;\\ \textbf{added}, changed, deleted, growth, total, percentage \\ of LOC in package in the past 3 months;\\ added, changed, deleted, growth, total, percentage \\ of LOC in package in the last 25 revisions;\end{tabular} \\ \hline
Warning history & \begin{tabular}[c]{@{}l@{}}warning modifications;\\ warning open revision;\\ \textbf{warning lifetime by revision}, by time;\end{tabular} \\ \hline
File characteristics & \begin{tabular}[c]{@{}l@{}}file type;\\ file name; \\package name;\end{tabular} \\ \hline
\end{tabular}
\label{tabel:variables}
\end{adjustbox}
\end{table}
\begin{table}[]
\caption{Number of Samples on Version 5.}
\small
\begin{adjustbox}{max width=0.48\textwidth}

\begin{tabular}{ll
>{\columncolor[HTML]{C0C0C0}}l l}
\hline
Project & Open/Unactionable & Close/Actionable & Delete \\ \hline
ant & 1061 & 54 & 0 \\
commons & 744 & 42 & 0 \\
tomcat & 1115 & 326 & 0 \\
jmeter & 468 & 145 & 7 \\
cass & 2245 & 356 & 64 \\
phoenix & 2046 & 343 & 13 \\
mvn & 790 & 28 & 44 \\
lucence & 2257 & 1168 & 440 \\
derby & 2386 & 121 & 0 \\ \hline
\end{tabular}
\end{adjustbox}
\label{table:numofSamples}
\end{table}


\subsection{Machine Learning Algorithms}

We choose three machine learning algorithms, i.e., Support Vector Machine (SVM), Random Forest (RF), Decision Tree (DT). These classifiers are selected for their common use in the software engineering literature. All these three algorithms are studied in Wang et al.'s paper~\citep{wang2018there} and the best performance is obtained by Random Forest, followed by Decision Tree. Regarding to SVM, it obtains the worst perform reported in six algorithms by Wang et al.~\citep{wang2018there}, but due to its wide combination with active learning and promising performance in many research areas like image retrieval~\citep{pasolli2013svm} and text classification~\citep{tong2001support}, especially imbalanced problems~\citep{ertekin2007active}, we also include this algorithm in our work. We now give a brief description of these algorithms and their application in this work.

All our learners come from the Python toolkit
Scikit-Learn~\citep{pedregosa2011scikit}. For the most part, we use the default parameters from that toolkit. Exception for support vector machines, we followed the advice of a previous publication~\citep{krishna2016bigse} 
which suggested using a linear, and a not radial, kernel).

\textbf{Support Vector Machine.} Support Vector Machine (SVM)~\citep{cortes1995support} is a supervised learning model for binary classification and regression analysis. The optimization objective of SVM is to maximize the margin, which is defined as the distance between the separating hyperplane (i.e., the decision boundary) and the training samples (i.e., support vectors) that are closest to the hyperplane. Support vector machine is a powerful linear model, it also can tackle nonlinear problems through the kernel trick, which consists of multiple hyperparameters that can be tuned to make good predictions.

\textbf{Random Forest.} Random forests~\citep{liaw2002classification} can be viewed as an ensemble of decision trees. The idea behind ensemble learning is to combine weak learners to build a more robust model or a strong learner, which has a better generalization error and is less susceptible to over-fitting. Such forests can be utilized for both classification and regression problems, and also employed to measure the relative importance of each feature on the prediction (by counting how often attributes are used in each tree of the forest).

\textbf{Decision Tree.} Decision tree learners are known for their ability to decompose complex decision processes into small and simple subsets~\citep{safavian1991survey}. In this process an associated multistage decision tree is hierarchically developed. There are several tree-based approaches widely used in software engineering areas like ID3, C4.5, CART and so forth. Decision tree is computationally cheap to use, and is easy for developers or managers to interpret.

\begin{algorithm}[!htbp]
\scriptsize
\SetKwInOut{Input}{Input}
\SetKwInOut{Output}{Output}
\SetKwInOut{Parameter}{Parameter}
\SetKwRepeat{Do}{do}{while}
\Input{$V_{n-1}$, previous version for training\\
$V_n$, current version for prediction\\ 
\textit{ C}, common set of features shared by five releases}
\Output{\textit{ Total Recall}, total recall for version n\\
\textit{ cost}, samples retrieved by percent}
\BlankLine


\BlankLine
\tcp{Keep reviewing until stopping rule satisfied}
\While{$|L_R| < 0.95|R|$}{
    \tcp{Start training or not}
    \eIf{$|L_R| \geq 1$}{
        $CL\leftarrow \mathit{Train}(L)$\;
        \tcp{Query next}
        $x\leftarrow \mathit{Query(CL},\neg L,L_R)$\;
    }{
        \tcp{Random Sampling}
        $x\leftarrow \mathit{Random}(\neg L)$\;
    }
    \tcp{Simulate review}
    $L_R,L\leftarrow \mathit{Include}(x,R,L_R,L)$\;
    $\neg L\leftarrow E \setminus L$\;
}
\Return{$L_R$}\;
\BlankLine
\Fn{Train($V_{n-1}$)}{
    \BlankLine
    \tcp{Classifier: Linear-SVM,decision tree, random forest}
    \BlankLine
    clf$\leftarrow \mathit{Classifier}$\;
    
    
    $ \mathit{training}_x, \mathit{training}_y \leftarrow V_{n-1} $\,
    
    \BlankLine
    clf$\leftarrow \mathit{clf.fit(training}_x, \mathit{training}_y)$\,
    
    \BlankLine
    \Return{clf}\;
}
\BlankLine
\Fn{PredictProb($V_n$,$ \mathit{clf}$)}{
    \BlankLine
    \tcp{predict Probability}
    \BlankLine
    $ \mathit{pos}_{\mathit{at}} \leftarrow \mathit{list(clf.classes).index("yes")}$\,
    
    $\mathit{testset}_x,\mathit{testset}_y \leftarrow V_n $\,
    
    $ \mathit{prob} \leftarrow \mathit{clf.PredictProb}(\mathit{testset}_x)[:, \mathit{pos}_{\mathit{at}}]$\,
    
    \BlankLine
    \Return{prob, $\mathit{testset}_y$}\;
}
\BlankLine
\Fn{Retrieve(prob, $\mathit{testset}_y $)}{
    \BlankLine
    \tcp{retrieve by descending-sorted probability }
    \BlankLine
    $ \mathit{sum} = 0$\,
    
    $ \mathit{order} \leftarrow \mathit{np}.\mathit{argsort(prob)}[::-1][:]$\,
    
    $ \mathit{pos}_{\mathit{all}} \leftarrow \mathit{number-of-positive- samples}$ \,
    
    $ \mathit{num}_{\mathit{all}} \leftarrow
    \mathit{length-of-testset}_y $\,
    
    \BlankLine
    \While{$ i \in \mathit{order} $}{
    \tcp{Sort label by descending order}
    $ \mathit{label}_{\mathit{real}} \leftarrow \mathit{testset}_\mathit{y[i]}$\,
    
    $ \mathit{sorted}_{\mathit{label}}.\mathit{append}(\mathit{label}_{\mathit{real}}) $\,
    
    \BlankLine
    
    \tcp{Retrieve}
    \While{$\mathit{label} \in \mathit{sorted}_{\mathit{label}} $}{
        \eIf{$\mathit{label} == "\mathit{yes}"$}{sum  $ +=1$\,
        
        $\mathit{pos}_{\mathit{get}} \leftarrow \mathit{sum}$\,}{continue}
    }
    $ \mathit{total}_{\mathit{recall}}.\mathit{append(pos}_{\mathit{get}} / \mathit{pos}_{\mathit{all}})$\,
    
     cost.append(len($\mathit{sorted}_{\mathit{label}}$) / $\mathit{num}_{\mathit{all}}$)
    
    }
    
    \Return{$\mathit{total}_{\mathit{recall}}$, cost}\;
}
\caption{Pseudo Code for Supervised Learning.}\label{alg:alg2}
\end{algorithm}

\section{Experiments}
\label{sec:evaluation}

In this section, we answer the four research questions formulated in Section \ref{sec:intro}.

\begin{RQ}
{\bf RQ1.} What is the baseline rate for bad static warnings?
\end{RQ}

\subsection{
Research method}


Static warning tools like FindBugs, Jlint and PMD are widely used in static warning analysis. Previous research has shown that FindBugs is more reliable than other SA tools regarding to its effective discrimination between true and false positives~\citep{wang2018there, rahman2014comparing}.
FindBugs is also known as a cost-efficient SA tool for detecting warnings by the combination of line-level, method-level and class-level granularity, thus reports much fewer warnings with obviously more lines ~\citep{rahman2014comparing, panichella2015would}.
Due to all the merits mentioned above, FindBugs has gained widespread popularity among individual users and technology-intensive companies, like Google\footnote{In 2009, Google held a global fixit for UMD's FindBugs tool and aimed at gathering feedback for the 4,000 highest confidence reported by FindBugs. It has been downloaded for more than a million times so far.}.

In terms of a baseline result, we used the default priority ranking reported by FindBugs. Since FindBugs generates warnings and classifies them into seven categories of patterns~\citep{shen2011efindbugs}, in which warnings with the same priority in random order have the same severity to be fixed. And the higher priority denotes that the warning report is more likely to be actionable suggested by FindBugs. This randomly ranking strategy provides a reasonable probabilistic bounded time for software developers to find bugs and implements the scene without any information to prioritize warning reports~\citep{heckman2011systematic,kremenek2004correlation}.


\subsection{Research results}

As is shown in Figure \ref{fig:results}, the dark blue dashed line denotes the learning curve of random selection generated from Findbugs reports. The curve grows diagonally, indicating that an end-user without any historical warning information or auxiliary tool has to inspect 2507 warnings to identify only 121 actionable ones in Derby dataset.

\begin{RQ}
{\bf RQ2.}  
What is the previous state-of-the-art method to tackle the prevalence of actionable warnings in SA tools?
\end{RQ}

\begin{table*}[]
\small
\caption{AUC \% on 9 projects for 10 runs. Our results are better than prior results (shown in blue) since they used default parameters in Weka while we adjusted (e.g.) the SVM kernel (as well as a more recent implementation of these tools).}
\tabcolsep=0.11cm
\begin{adjustbox}{max width=1\textwidth}
\begin{tabular}{@{}lllllllllllll@{}}
\toprule
\multicolumn{1}{c}{} & \multicolumn{2}{c}{Active+SVM} & \multicolumn{2}{c}{Supervised\_SVM} & \multicolumn{2}{c}{Active+RF} & \multicolumn{2}{c}{Supervised\_RF} & \multicolumn{2}{c}{Active+DT} & \multicolumn{2}{c}{Supervised\_DT} \\ \cmidrule(l){2-13} 
\multicolumn{1}{c}{\multirow{-2}{*}{\textbf{Project}}} & \textbf{Median} & IQR & \textbf{\begin{tabular}[c]{@{}l@{}}Median\\ (IQR)\end{tabular}} & \begin{tabular}[c]{@{}l@{}}Median of\\ Prior work\end{tabular} & \textbf{Median} & IQR & \textbf{\begin{tabular}[c]{@{}l@{}}Median\\ (IQR)\end{tabular}} & \begin{tabular}[c]{@{}l@{}}Median of\\ Prior work\end{tabular} & \textbf{Median} & IQR & \textbf{\begin{tabular}[c]{@{}l@{}}Median\\ (IQR)\end{tabular}} & \begin{tabular}[c]{@{}l@{}}Median of\\ Prior work\end{tabular} \\ \cmidrule(r){1-1}
Derby & \cellcolor[HTML]{FFCCC9}98 & \cellcolor[HTML]{FFCCC9}1 & \cellcolor[HTML]{FFCCC9}97(2) & \cellcolor[HTML]{CBCEFB}50 & \cellcolor[HTML]{FFCCC9}{\color[HTML]{333333} 96} & \cellcolor[HTML]{FFCCC9}{\color[HTML]{333333} 7} & \cellcolor[HTML]{FFCCC9}{\color[HTML]{333333} 97(4)} & \cellcolor[HTML]{CBCEFB}{\color[HTML]{333333} 43} & \cellcolor[HTML]{FFCCC9}93 & \cellcolor[HTML]{FFCCC9}2 & \cellcolor[HTML]{FFCCC9}94(4) & \cellcolor[HTML]{CBCEFB}44 \\
Mvn & \cellcolor[HTML]{FFCCC9}94 & \cellcolor[HTML]{FFCCC9}3 & \cellcolor[HTML]{FFCCC9}96(7) & \cellcolor[HTML]{CBCEFB}50 & \cellcolor[HTML]{FFCCC9}93 & \cellcolor[HTML]{FFCCC9}2 & \cellcolor[HTML]{FFCCC9}97(3) & \cellcolor[HTML]{CBCEFB}45 & 67 & 3 & 91(2) & \cellcolor[HTML]{CBCEFB}45 \\
Lucence & \cellcolor[HTML]{FFCCC9}95 & \cellcolor[HTML]{FFCCC9}1 & \cellcolor[HTML]{FFCCC9}97(3) & \cellcolor[HTML]{CBCEFB}50 & 85 & 9 & 99(2) & \cellcolor[HTML]{CBCEFB}98 & \cellcolor[HTML]{FFCCC9}94 & \cellcolor[HTML]{FFCCC9}2 & \cellcolor[HTML]{FFCCC9}93(4) & \cellcolor[HTML]{CBCEFB}98 \\
Phoenix & \cellcolor[HTML]{FFCCC9}97 & \cellcolor[HTML]{FFCCC9}2 & \cellcolor[HTML]{FFCCC9}97(3) & \cellcolor[HTML]{CBCEFB}62 & 90 & 7 & 97(3) & \cellcolor[HTML]{CBCEFB}71 & \cellcolor[HTML]{FFCCC9}90 & \cellcolor[HTML]{FFCCC9}2 & \cellcolor[HTML]{FFCCC9}91(7) & \cellcolor[HTML]{CBCEFB}70 \\
Cass & \cellcolor[HTML]{FFCCC9}96 & \cellcolor[HTML]{FFCCC9}5 & \cellcolor[HTML]{FFCCC9}99(3) & \cellcolor[HTML]{CBCEFB}67 & \cellcolor[HTML]{FFCCC9}96 & \cellcolor[HTML]{FFCCC9}4 & \cellcolor[HTML]{FFCCC9}98(5) & \cellcolor[HTML]{CBCEFB}70 & \cellcolor[HTML]{FFCCC9}90 & \cellcolor[HTML]{FFCCC9}1 & \cellcolor[HTML]{FFCCC9}94(4) & \cellcolor[HTML]{CBCEFB}69 \\
Jmeter & \cellcolor[HTML]{FFCCC9}94 & \cellcolor[HTML]{FFCCC9}1 & \cellcolor[HTML]{FFCCC9}95(2) & \cellcolor[HTML]{CBCEFB}50 & 90 & 4 & 97(2) & \cellcolor[HTML]{CBCEFB}86 & \cellcolor[HTML]{FFCCC9}86 & \cellcolor[HTML]{FFCCC9}2 & \cellcolor[HTML]{FFCCC9}91(12) & \cellcolor[HTML]{CBCEFB}82 \\
Tomcat & \cellcolor[HTML]{FFCCC9}98 & \cellcolor[HTML]{FFCCC9}1 & \cellcolor[HTML]{FFCCC9}97(3) & \cellcolor[HTML]{CBCEFB}50 & \cellcolor[HTML]{FFCCC9}92 & \cellcolor[HTML]{FFCCC9}5 & \cellcolor[HTML]{FFCCC9}96(2) & \cellcolor[HTML]{CBCEFB}80 & \cellcolor[HTML]{FFCCC9}94 & \cellcolor[HTML]{FFCCC9}2 & \cellcolor[HTML]{FFCCC9}92(6) & \cellcolor[HTML]{CBCEFB}64 \\
Ant & \cellcolor[HTML]{FFCCC9}95 & \cellcolor[HTML]{FFCCC9}2 & \cellcolor[HTML]{FFCCC9}98(2) & \cellcolor[HTML]{CBCEFB}50 & \cellcolor[HTML]{FFCCC9}94 & \cellcolor[HTML]{FFCCC9}1 & \cellcolor[HTML]{FFCCC9}98(3) & \cellcolor[HTML]{CBCEFB}44 & 84 & 3 & 94(7) & \cellcolor[HTML]{CBCEFB}44 \\
Commons & 91 & 3 & 98(3) & \cellcolor[HTML]{CBCEFB}50 & \cellcolor[HTML]{FFCCC9}93 & \cellcolor[HTML]{FFCCC9}1 & \cellcolor[HTML]{FFCCC9}92(2) & \cellcolor[HTML]{CBCEFB}57 & \cellcolor[HTML]{FFCCC9}80 & \cellcolor[HTML]{FFCCC9}8 & \cellcolor[HTML]{FFCCC9}85(14) & \cellcolor[HTML]{CBCEFB}56 \\ \bottomrule
\end{tabular}
\label{table:AUC}
\end{adjustbox}
\end{table*}

\subsection{Research method}

Wang et al.~\citep{wang2018there} implements a Java tool to extract the value of 116 total features collected from exhaustive systematic literature review and employs the machine learning utility Weka\footnote{https://www.cs.waikato.ac.nz/~ml/weka/} to build classification models. An optimal SA feature set with 23 features is identified as the golden features by obtaining the best AUC values evaluated with 6 machine learning classifiers.
We reproduce the experiments with three most outperforming supervised learning models in the previous research study, e.g., weighted linear SVM, random forest and decision tree with default parameters in Python3.7.
The detailed process to replicate the baseline is demonstrated in Algorithm~\ref{alg:alg2}.


The specific process is as follows:
For each project, a supervised model (either weighted SVM, Random Forest or Decision Tree) is built by training on Version 4. After the training process, we test on Version 5 for the same project and get a list of probability for each bug reported by FindBugs to be actionable. Sort this list of probability from most likely to be real actionable to least likely and retrieve these warnings in a descending order to report the \textit{total recall}, \textit{cost} and \textit{AUC} as evaluation metrics.

\subsection{Research results}

As shown in Table~\ref{table:AUC}, the median and IQR of AUC scores of ten runs on nine projects are reported in our paper. \textit{Median} and \textit{IQR} are commonly used robust measures of a set of observations. \textit{IQR}~(the interquartile range) is a measure of statistical dispersion. It evaluates the variability of distribution by dividing a data set into quartiles and reflecting the difference between 75th and 25th percentiles.

For three supervised learning methods explored, Linear weighted Support Vector Machine and Random Forest both outperform Decision Tree. For incremental active learning algorithms, the best combination is \textit{Active Learning + Support Vector Machine}, followed by \textit{Active Learning + Random Forest} and \textit{Active Learning + Decision Tree}. 

It's observed that incremental active learning can obtain high AUC, no worse than supervised learning on most of datasets. The pink shadow highlights the median results of active learning methods which are better or no less 0.05 than the median AUC of the state-of-the-art methods.

The column "Prior Work" shows results reported in Wang et al.'s prior research~\citep{wang2018there}.
Note that our AUC scores for supervised models replicated with Python3.7 are higher than that prior work implemented by Weka. This difference is explained by two factors.
\begin{itemize}
\item
We found that better results could be obtained by adjusting some of the learner parameters; e.g. we use a linear (not radial) kernel for our SVM.
\item
The implementation tools employed by our study and previous work are different. Prior work used a Java implementation of these tools (in Weka) while our replication utilizes a more recent Python toolkit (Scikit-Learn) that is being used and updated by a larger and more developed community.
\end{itemize}

\begin{RQ}
{\bf RQ3.} Does incremental active learning reduce the cost to identify actionable Static Warnings?
\end{RQ}

The purpose of this research question is to compare incremental active learning with random selection and traditional supervised learning models.

\subsection{Research method
}

Considering a real-world scenario when a software project in different stages of the life cycle,
RQ3 is answered in two parts: We first contrast incremental active learning, denoted as solid lines in Figure~\ref{fig:results} with random ranking (default ranking reported from FindBugs, denoted as dark blue dashed line in Figure~\ref{fig:results}). Then, we compare active learning results with supervised learning (denoted as purple, lighted blue and red dashed lines in Figure~\ref{fig:results}).


\subsection{Research results}

\begin{figure*}[!htbp]
    \centering
    \subfigure[commons]{\includegraphics[width=0.32\textwidth, trim = {0.1cm 0.1cm 2.8cm 0.5cm}, clip]{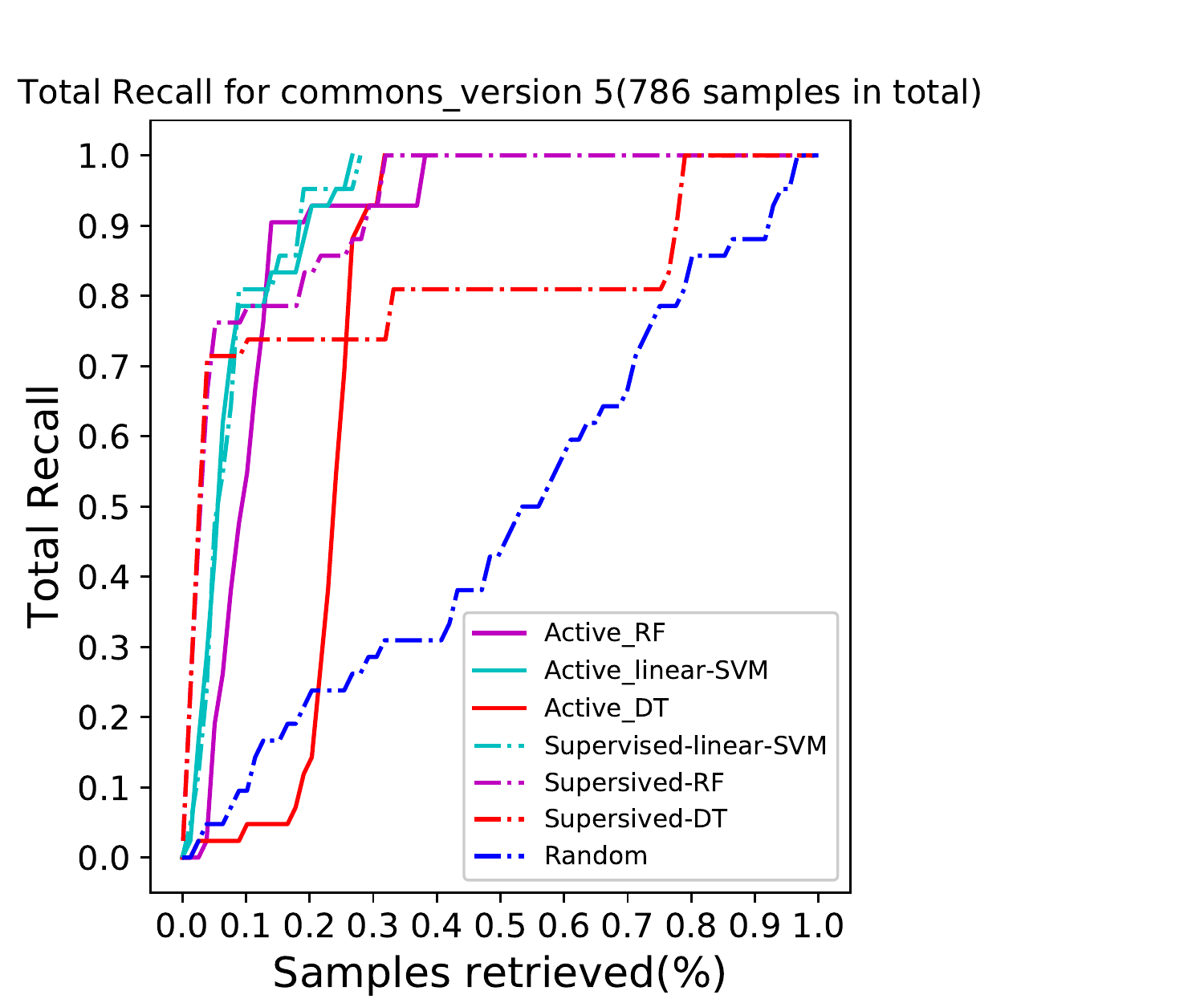}} 
    \subfigure[tomcat]{\includegraphics[width=0.32\textwidth, trim = {0.1cm 0.1cm 2.8cm 0.5cm}, clip]{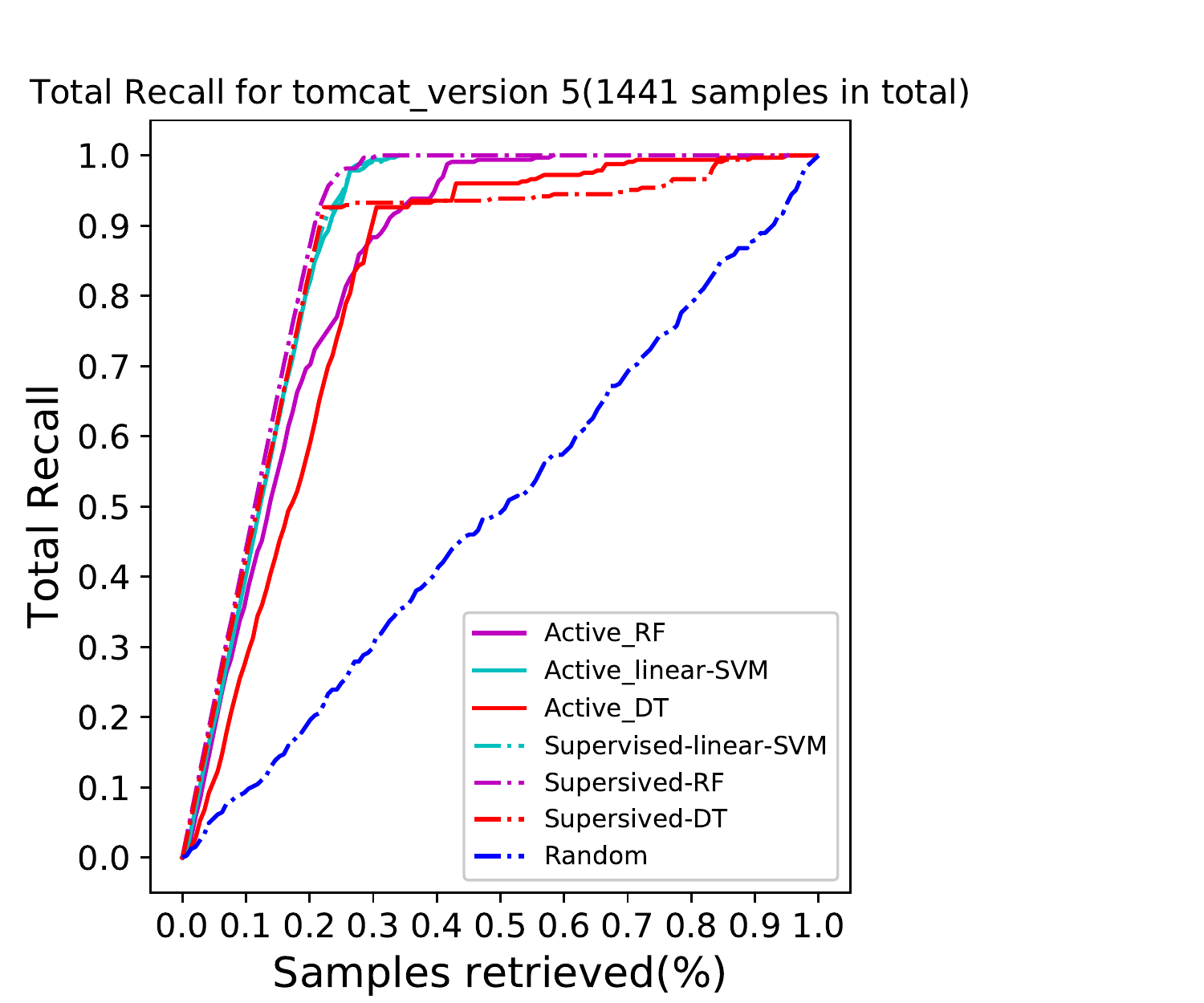}}
    \subfigure[jmeter]{\includegraphics[width=0.32\textwidth, trim = {0.1cm 0.1cm 2.8cm 0.5cm}, clip]{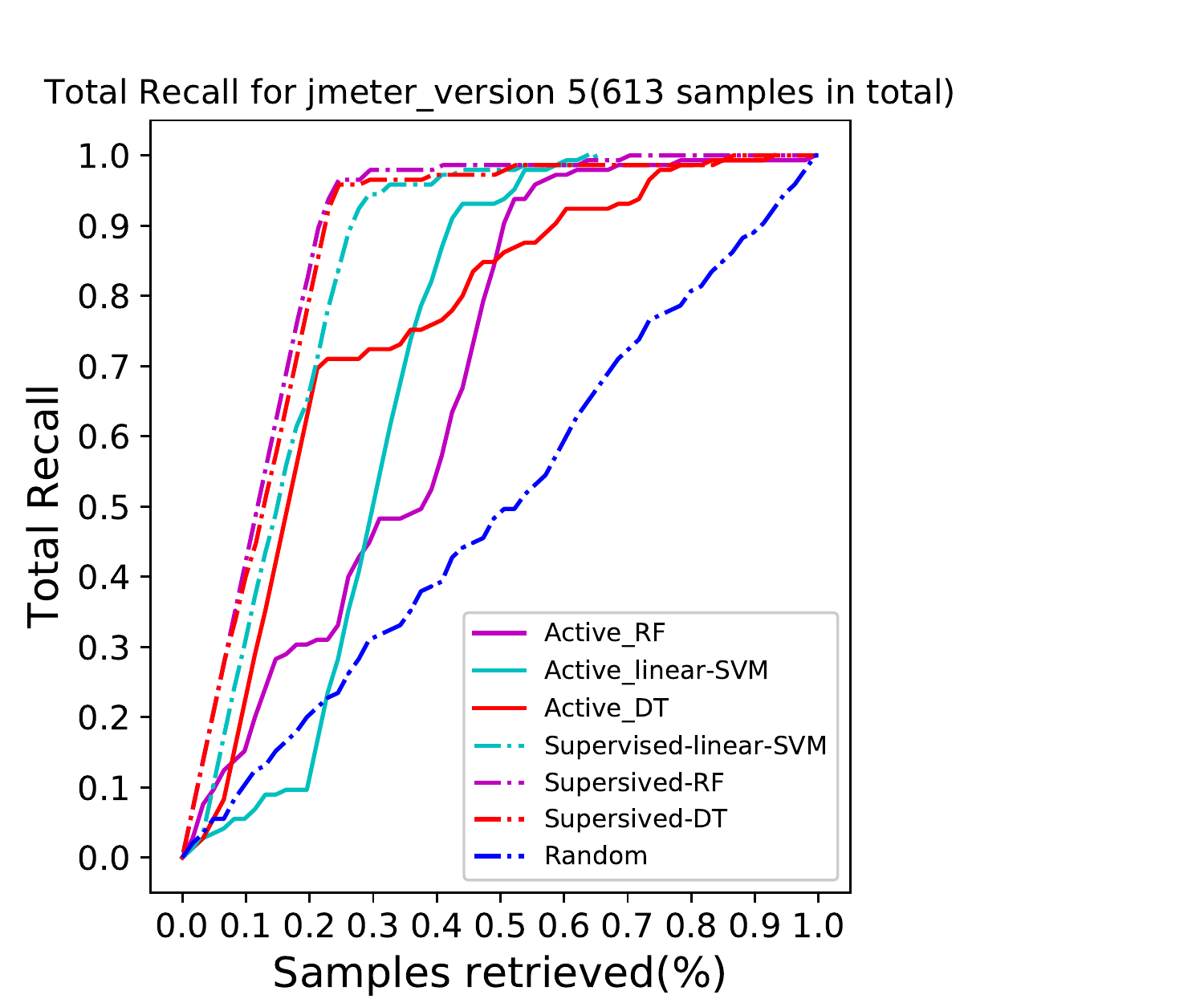}}
    \subfigure[cass]{\includegraphics[width=0.32\textwidth, trim = {0.1cm 0.1cm 2.8cm 0.5cm}, clip]{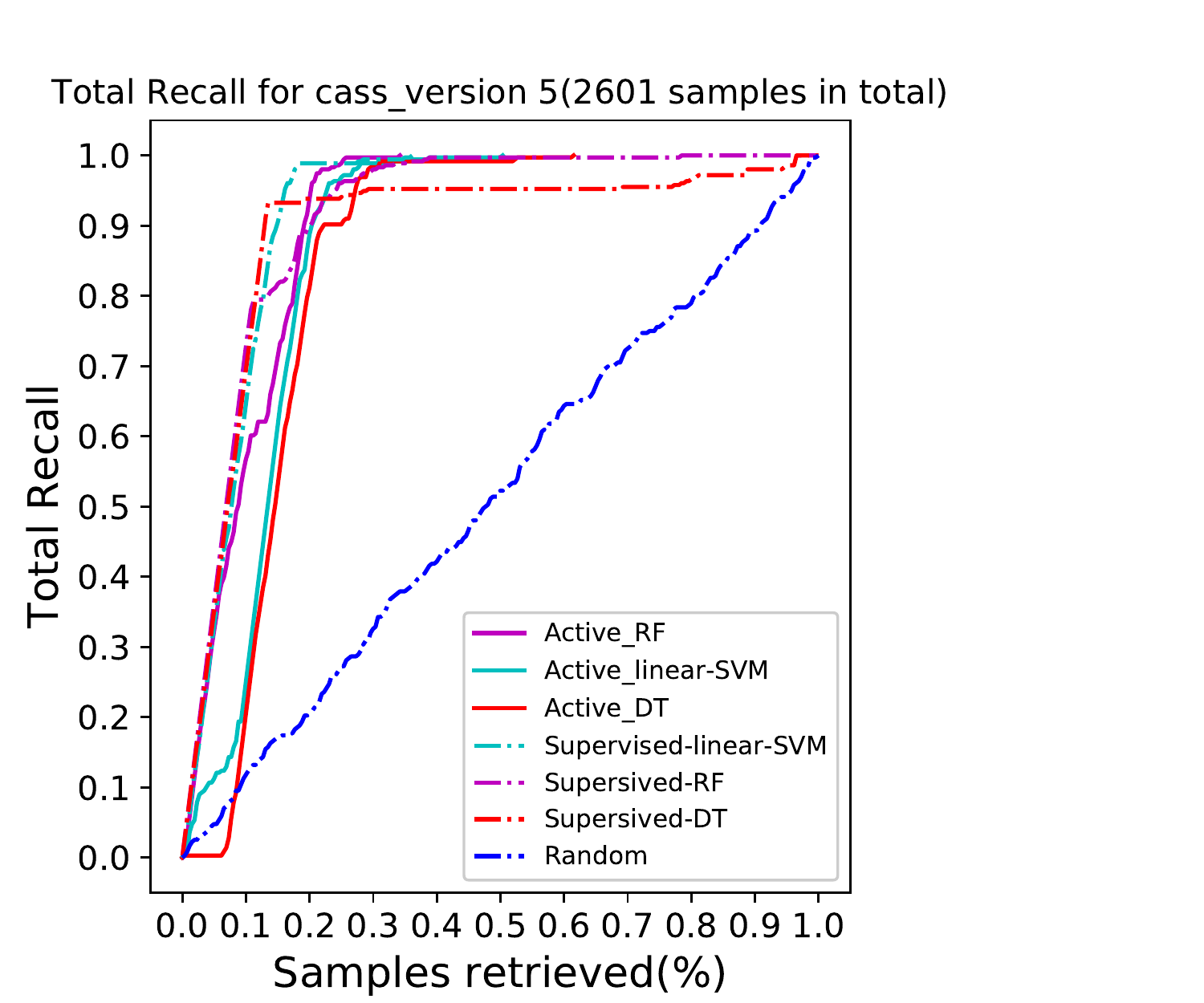}}
    \subfigure[derby]{\includegraphics[width=0.32\textwidth, trim = {0.1cm 0.1cm 2.8cm 0.5cm}, clip]{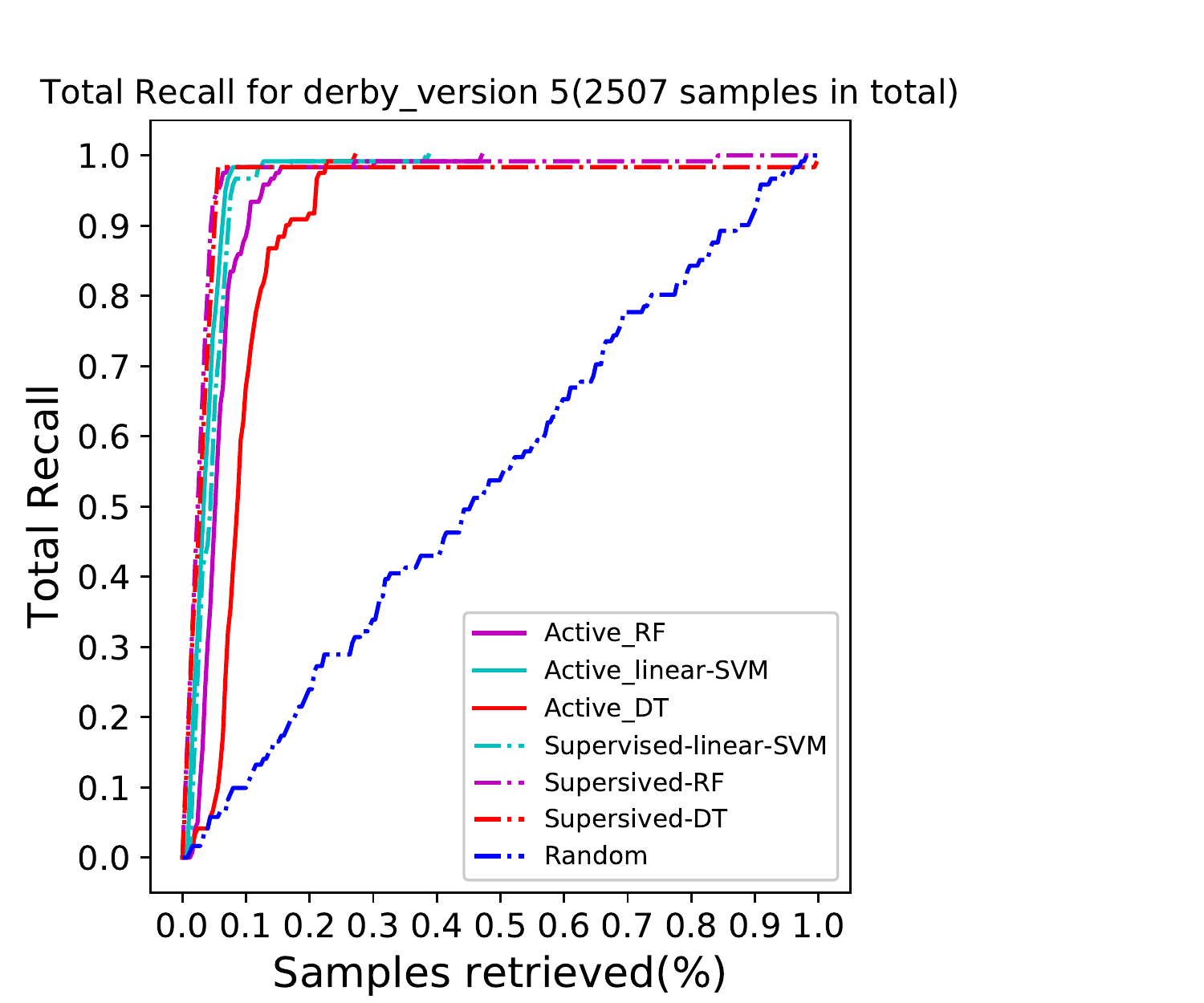}}
    \subfigure[phoenix]{\includegraphics[width=0.32\textwidth, trim = {0.1cm 0.1cm 2.8cm 0.5cm}, clip]{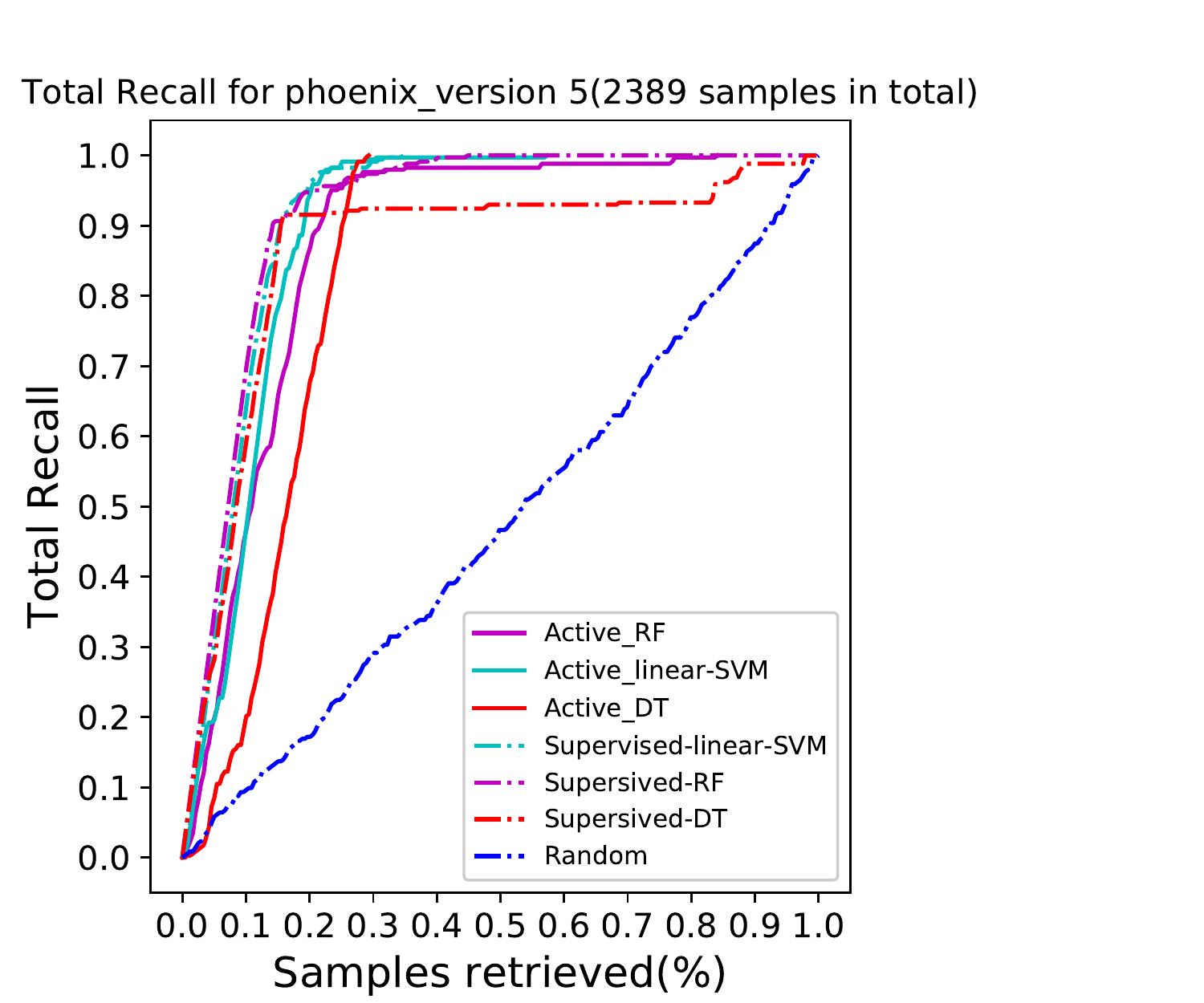}}
    \subfigure[lucence]{\includegraphics[width=0.32\textwidth, trim = {0.1cm 0.1cm 2.8cm 0.5cm}, clip]{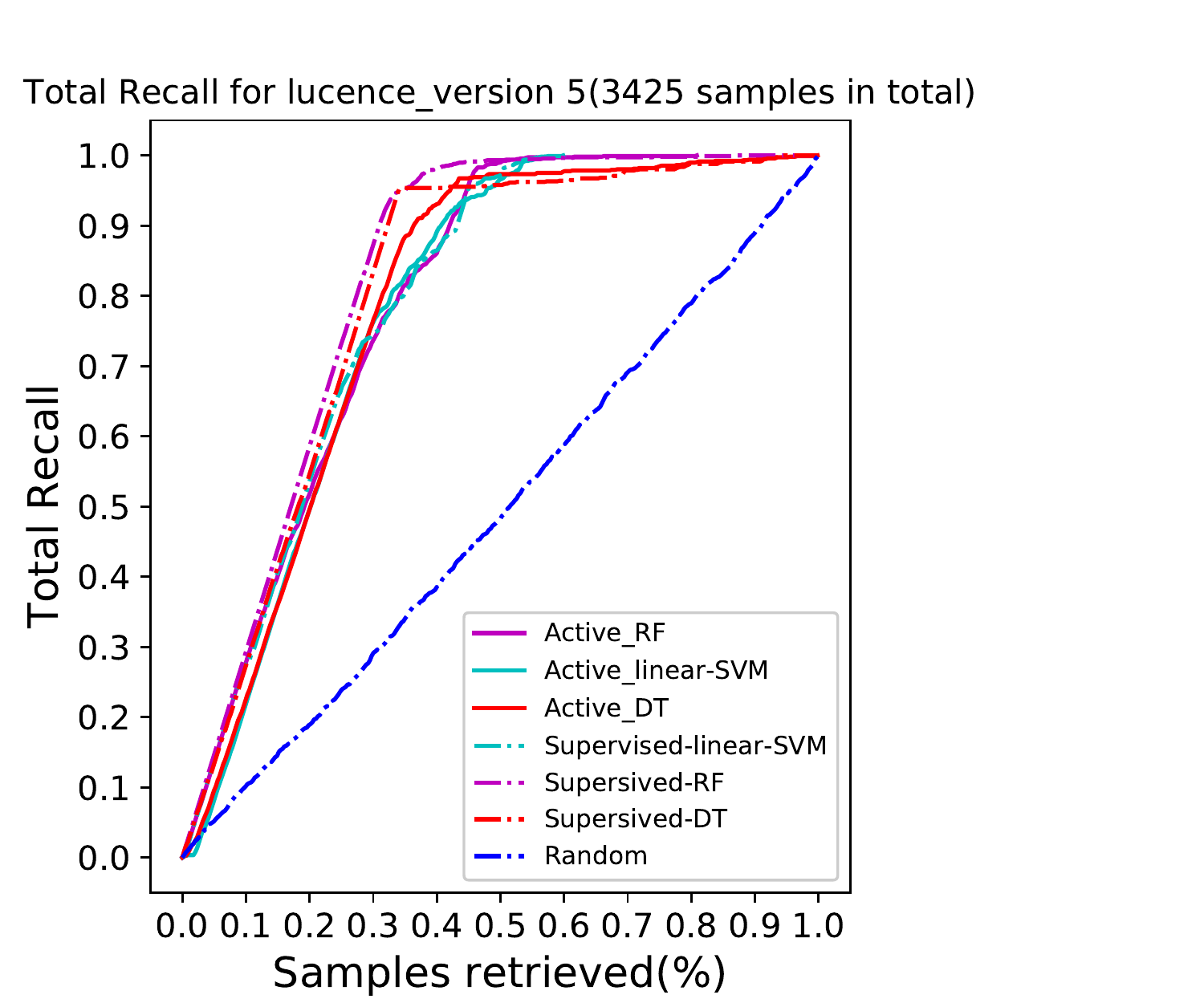}}
    \subfigure[mvn]{\includegraphics[width=0.32\textwidth, trim = {0.1cm 0.1cm 2.8cm 0.5cm}, clip]{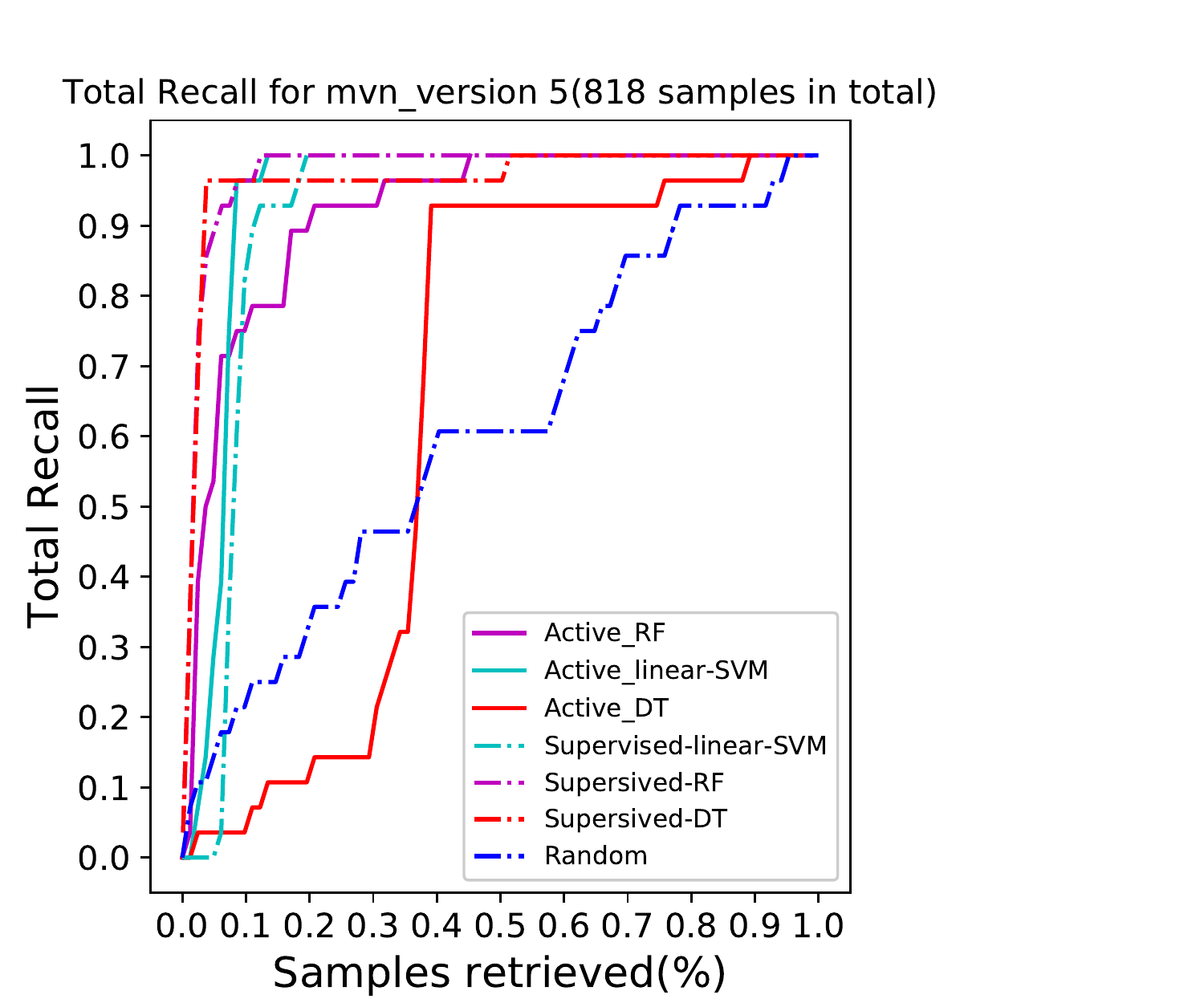}}
    \subfigure[ant]{\includegraphics[width=0.32\textwidth, trim = {0.1cm 0.1cm 2.8cm 0.5cm}, clip]{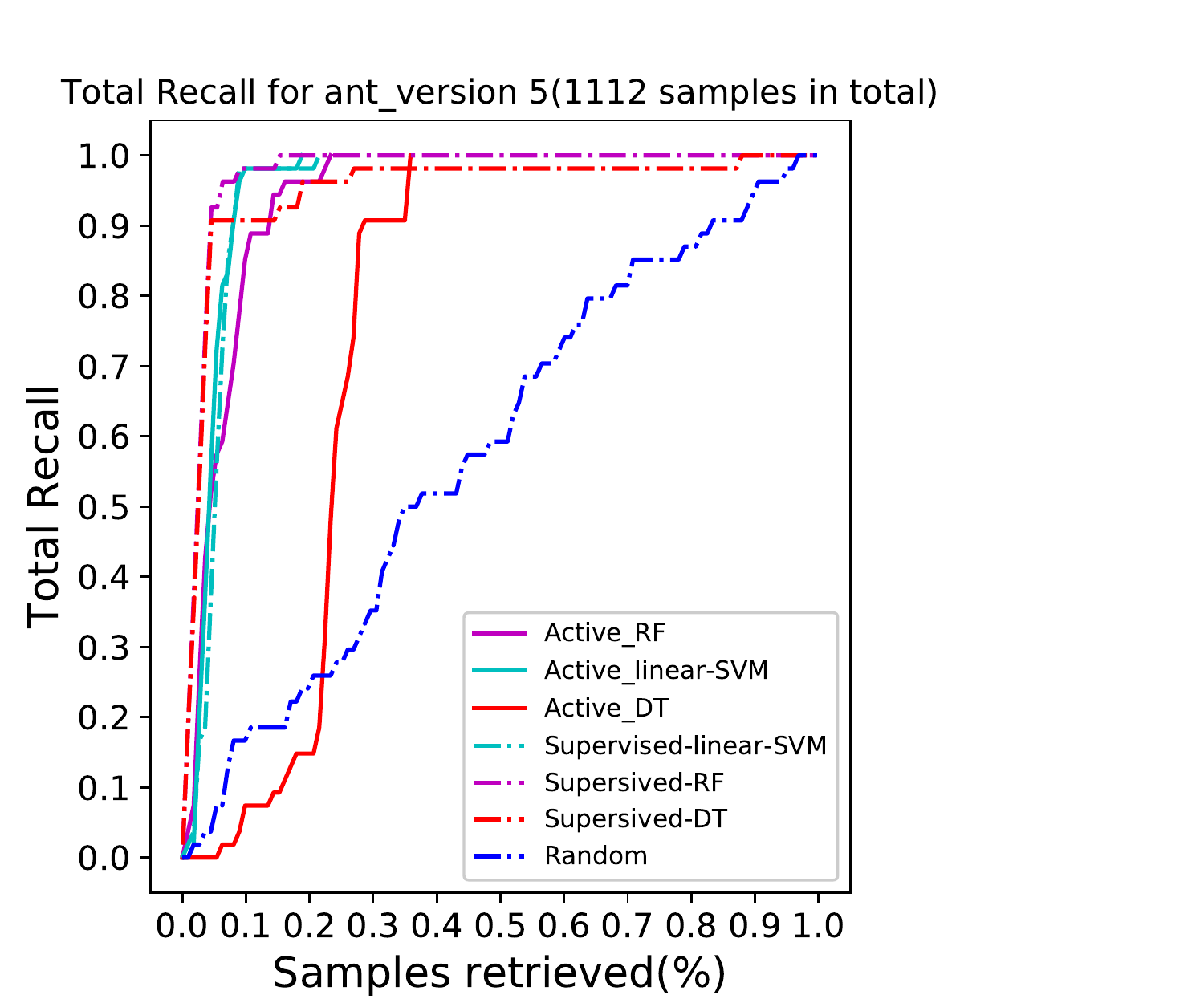}}
    \caption{Test Results of incremental active learning, supervised learning and randomly selection.}
    \label{fig:results}
\end{figure*}

Results of supervised learning methods are denoted as light blue, purple and red dashed lines. As revealed in Figure~\ref{fig:results}, Random Forest outperforms the other classifiers, followed by Linear SVM and Decision Tree.

Figure \ref{fig:results} provides an overall view of the experiment results to address Research Question 3. These nine subplots are the results of a ten-time repeated experiment on fourth and fifth versions of nine projects and we only report the median values here. The latest version 5 is selected to construct incremental active learning, while for the supervised learning model, we choose the two latest versions, learning patterns from version fourth for model construction and testing on version fifth for evaluation to make the experimental results comparable.

\begin{figure*}[!htbp]
    \centering
    \subfigure{\includegraphics[width=0.32\textwidth]{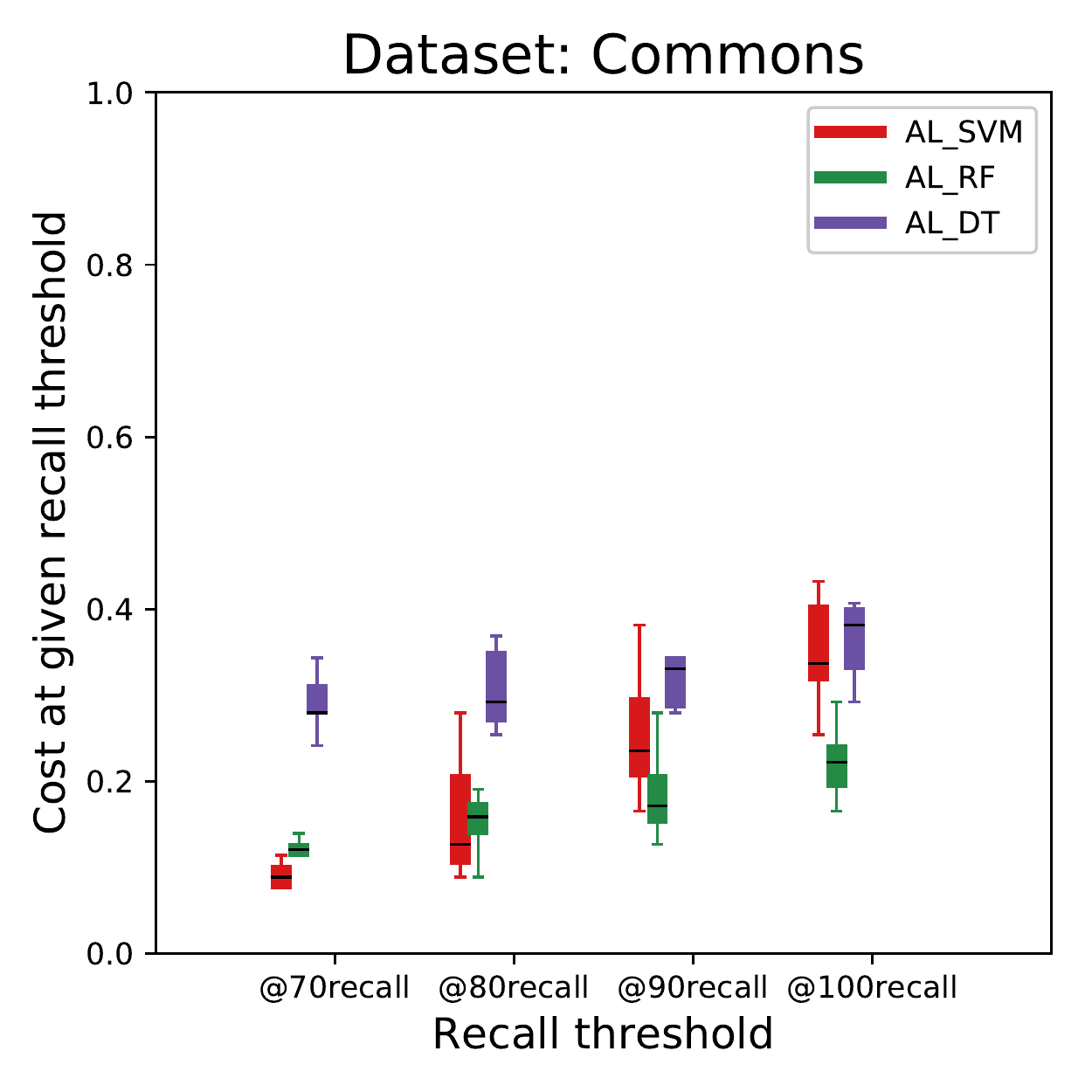}} 
    \subfigure{\includegraphics[width=0.32\textwidth]{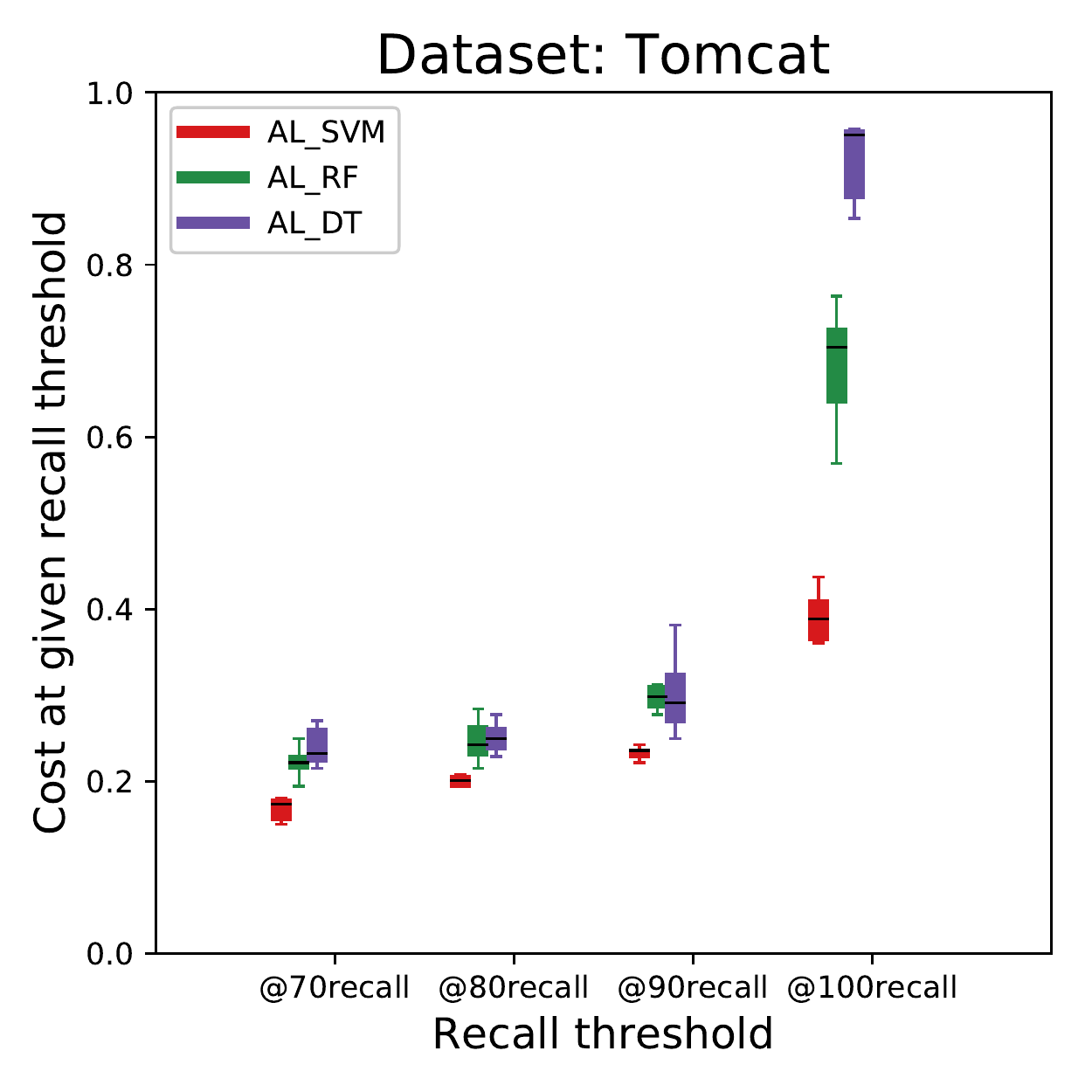}}
    \subfigure{\includegraphics[width=0.32\textwidth]{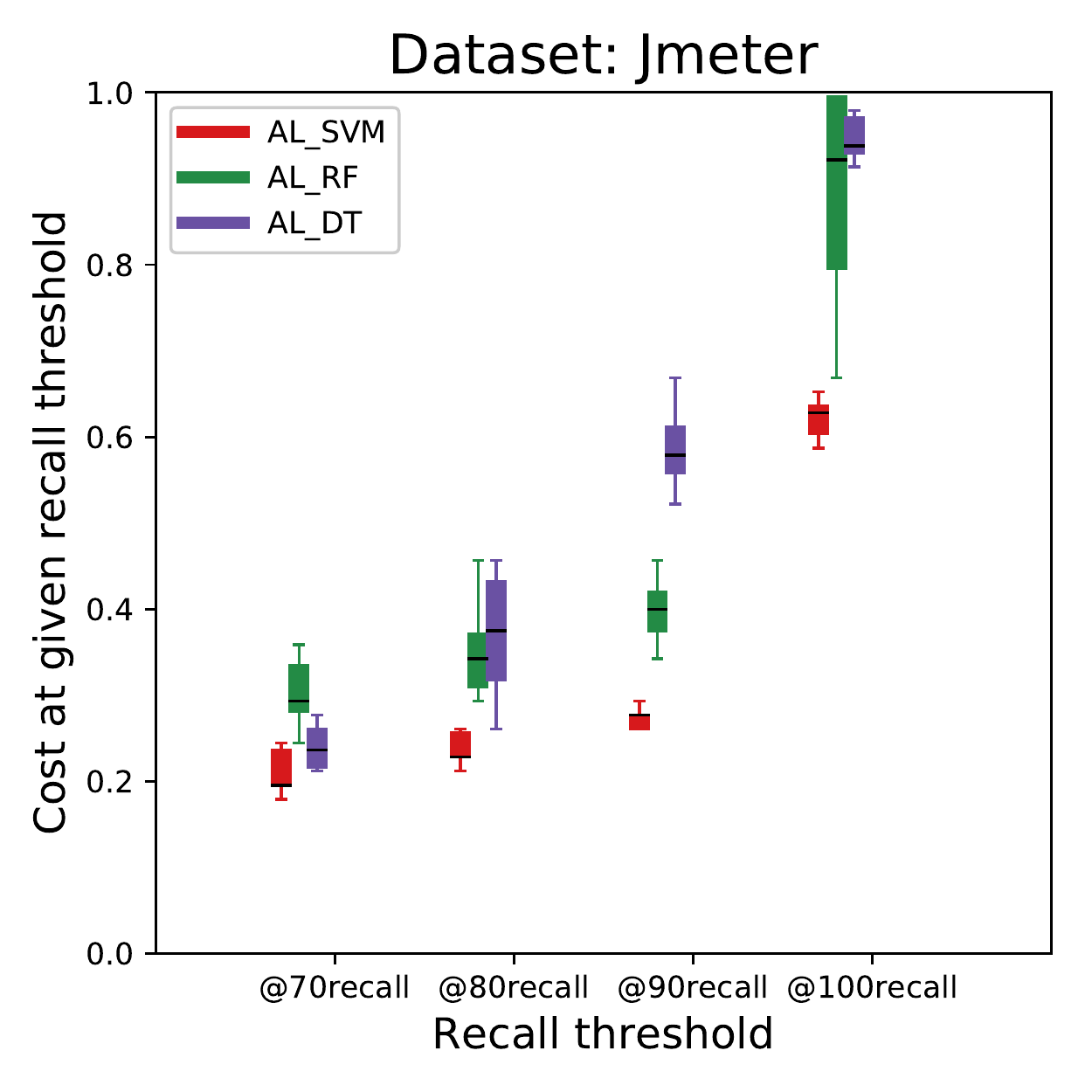}}
    \subfigure{\includegraphics[width=0.32\textwidth]{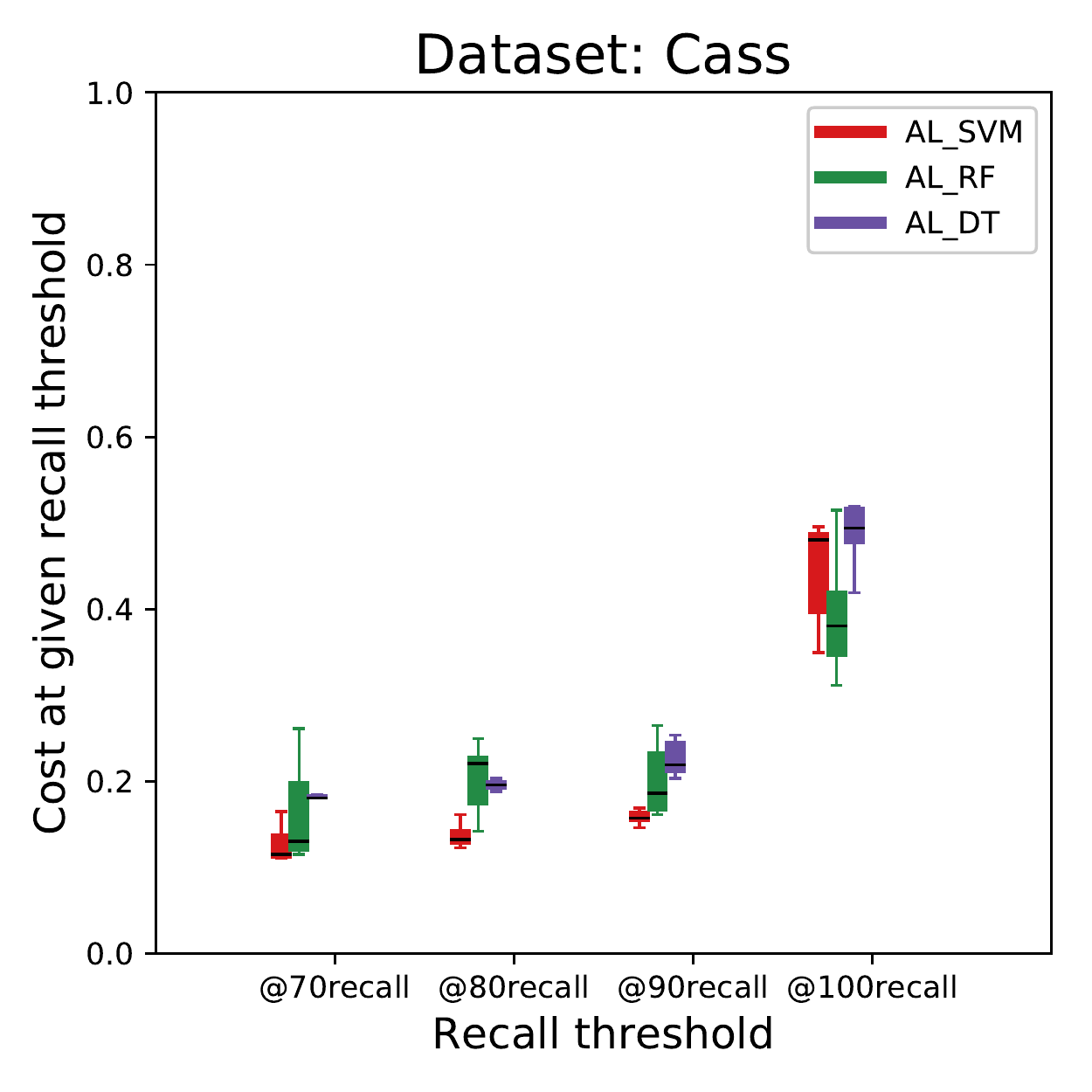}}
    \subfigure{\includegraphics[width=0.32\textwidth]{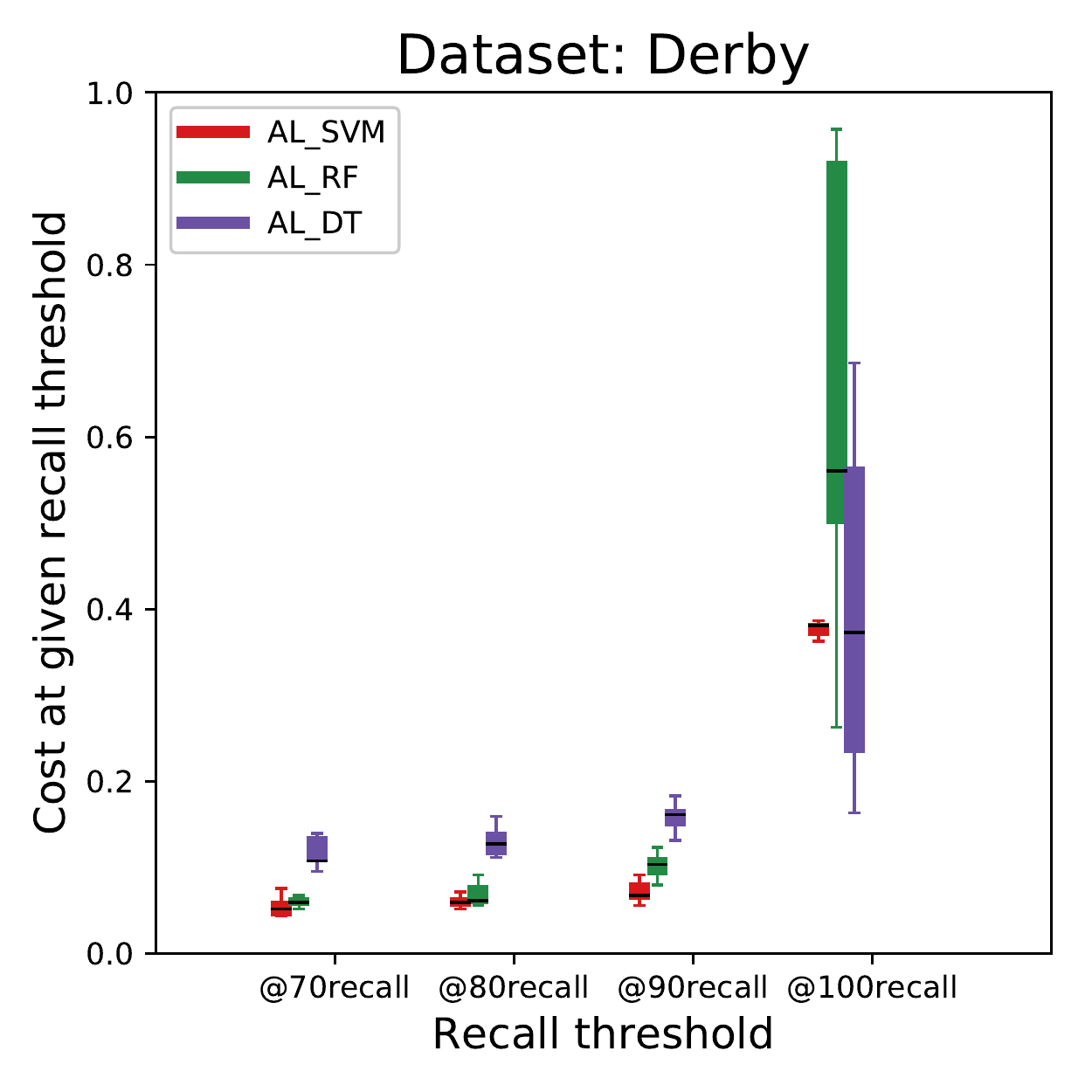}}
    \subfigure{\includegraphics[width=0.32\textwidth]{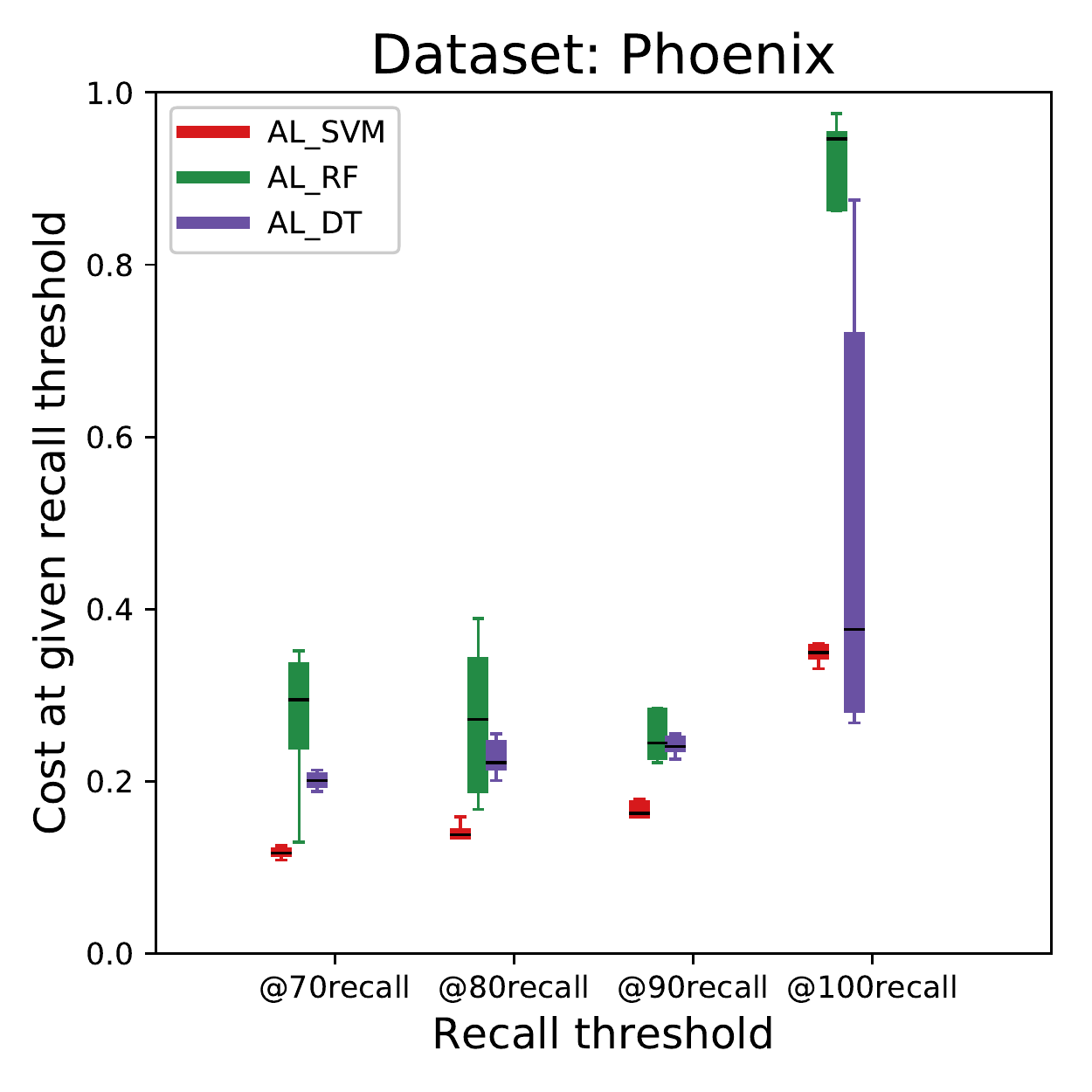}}
    \subfigure{\includegraphics[width=0.32\textwidth]{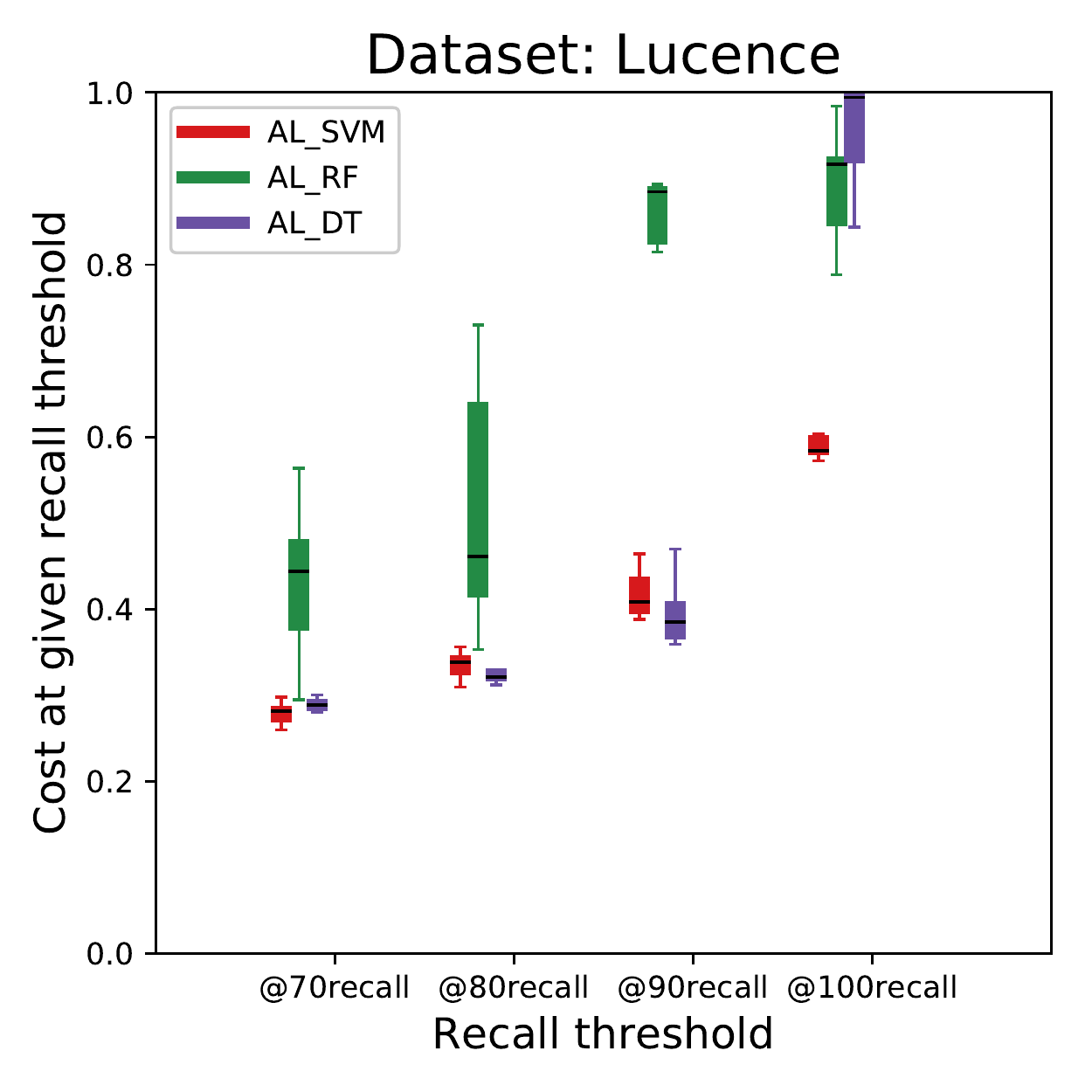}}
    \subfigure{\includegraphics[width=0.32\textwidth]{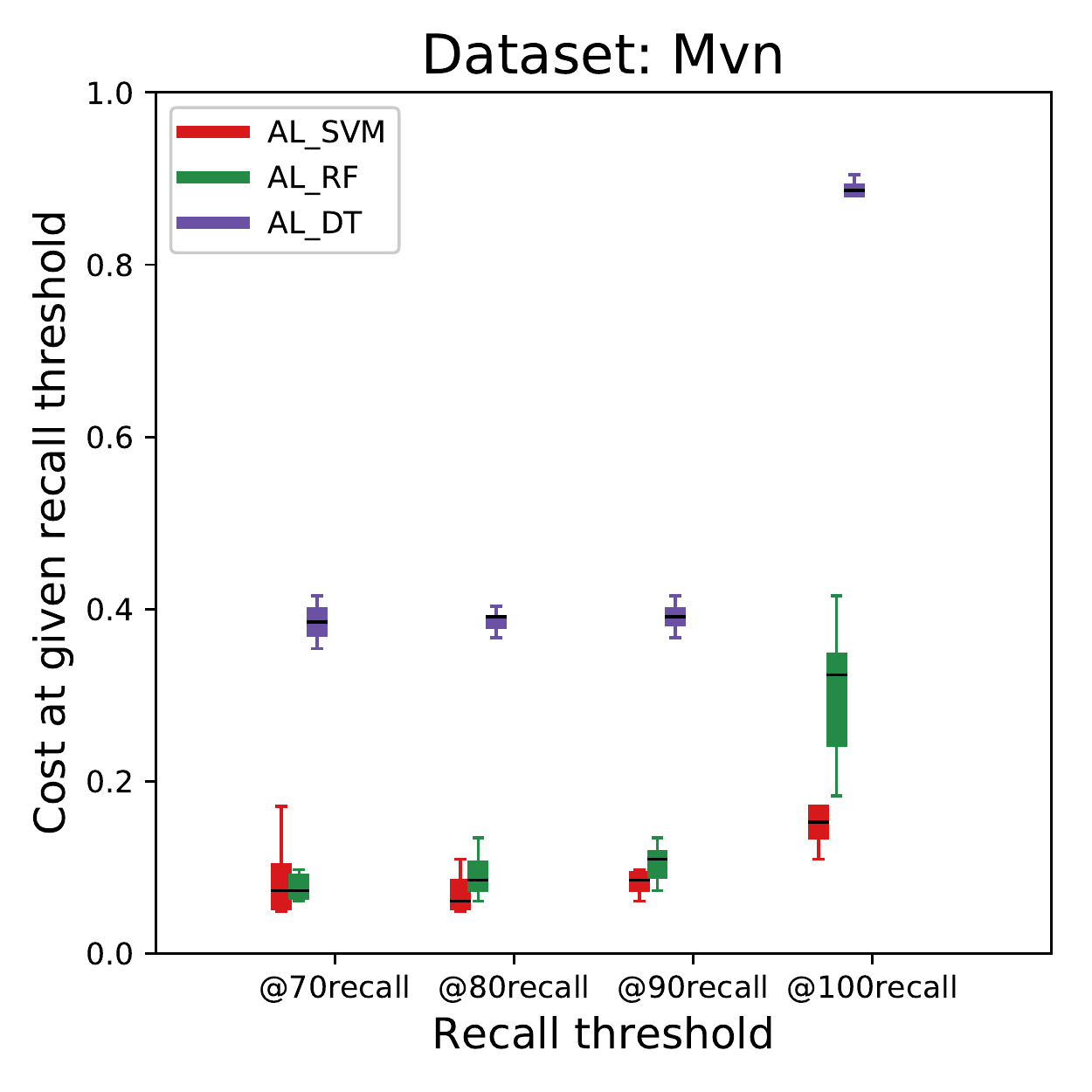}}
    \subfigure{\includegraphics[width=0.32\textwidth]{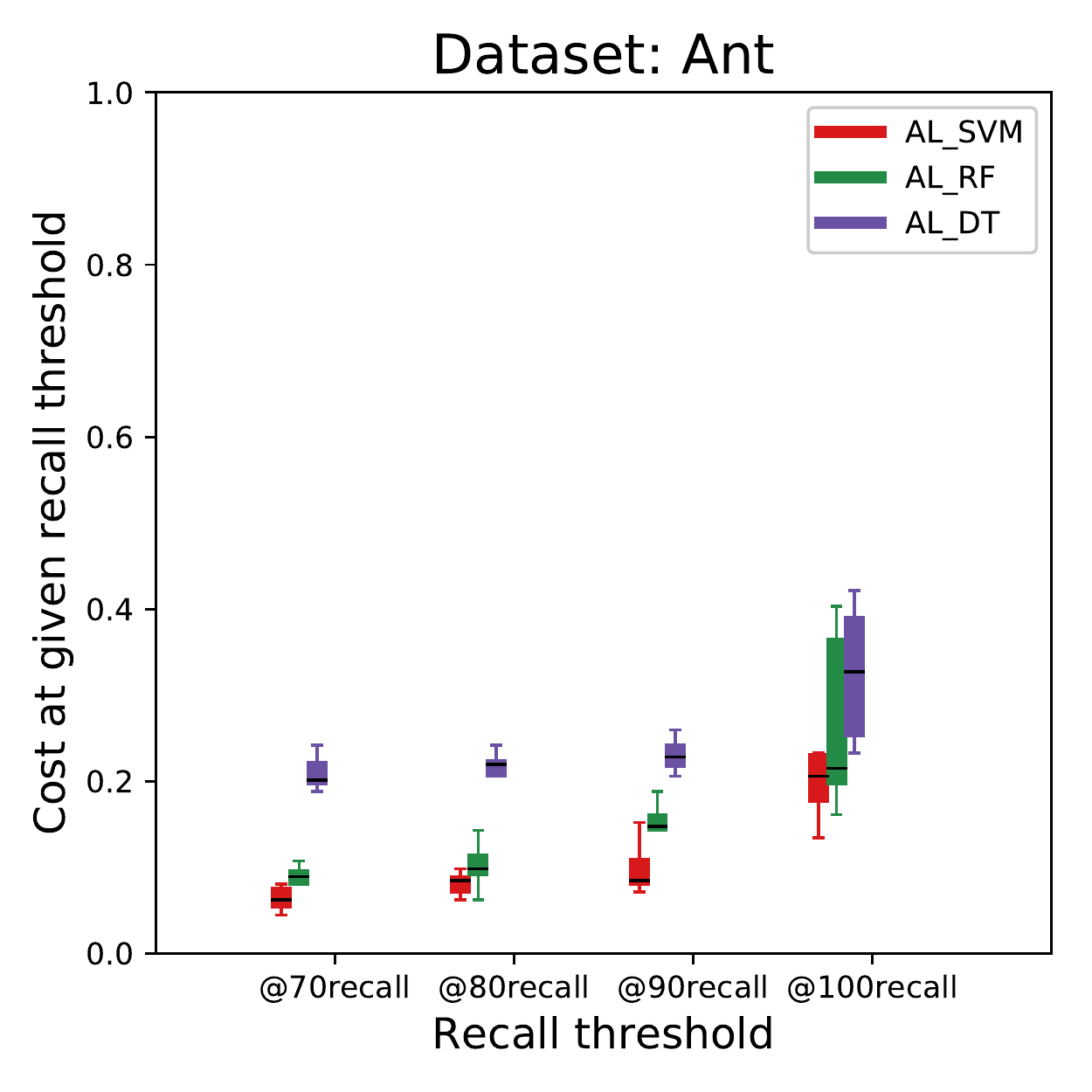}}
    \caption{Cost Results at different thresholds for Incremental Active Learning. }
    \label{fig:boxplots}
\end{figure*}

Figure \ref{fig:boxplots} summarizes the ratio of real actionable warnings in version 5 of each project and the corresponding median of cost when applying incremental active learning to identify all these actionable warnings. 

As illustrated in Figure \ref{fig:results}, incremental active learning outperforms random selection, which simulates real-time cost bound when an end-user recurs to warning reports prioritized by FindBugs.
While, the learning curve of incremental active learning without historical version is almost as good as supervised learning in most of nine projects based on version history. Also, the test results on nine datasets suggest that \textit{Linear SVM $+$ incremental active learning} is the best combination of all active learning schemes, and \textit{Random Forest} is the winner in supervised learning methods.

Overall, the above observations suggest that applying an incremental active learning model in static warning identification can help to retrieve actionable warnings in higher priority and reduce the effort to eliminate false alarms for software projects without adequate version history.

\begin{RQ}
{\bf RQ4.} How many samples should be retrieved to identify all the actionable Static Warnings?
\end{RQ}

How many samples to be retrieved is a critical problem when implementing an active learning model in the scenario of static warning identification. Stopping too early or too late will incur the issue of missing important actionable warnings or wasting unnecessary running time and CPU resources. 

In the following part, we introduce the research method and analysis of the experimental results to answer Research Question 4.

\subsection{Research method}

Figure \ref{fig:boxplots} employs the box-plot to describe the costs required or percentage of samples retrieved by three classifiers, Linear weighted SVM, Random Forest and Decision Tree combined with our incremental active learning algorithm. 
Horizontal coordinate of the box charts represents the thresholds of recall, a mechanism to stop retrieving new potential actionable warnings when the proportion of related samples found reached the specific given thresholds. The vertical axis means the corresponding effort required to obtain the given recall, measured by the proportion of warnings retrieved.

\subsection{Research results}

Based on the results shown in Figure \ref{fig:boxplots}, it can be observed that the growth of effort required is in a gentle and slow fashion when the threshold of relevant warnings visited increasing from 70 \% to 90 \%.
However, for reaching 100 \% threshold, the effort needed is almost or over twice compared with the cost of threshold equal to 90 \%. A very intuitive suggestion can be obtained from Figure \ref{fig:boxplots} is learning from 20 \% or 30 \% warnings for each of these nine projects, in which case the active learning models can identify over 90 \% of actionable warnings.

However, there is an exception. Results of lucence reveal that our model has to inspect more than 40 \% of data to identify 90 \% actionable warnings. Revisiting Table \ref{table:numofSamples}, it indicts that most of our projects have data imbalanced issues (ratio of target class is less than 20 \% for derby, mvn, phoenix, cass, commons and ant, and for jmeter and tomcat it's slightly over 20 \%) while ratio of lucence (about 35 \%) is relatively higher. Our study attempts to provide a solid guideline but there is no general conclusion about the specific percent of data that should be fed into the learner. It highly depends on the degree of data imbalance and the trade-off between missing target samples and reducing costs. Since the cost can only be reduced at the expense of a lower threshold, which means missing some real actionable warnings.

In summary,
our model has been proven to be an efficient methodology to deal with information retrieval problem for SA identification of extremely unbalanced data sets, moreover, it is also a good option for engineers and researchers to apply active learning model in general problems because it has a lower building cost, a wider application range, and a higher efficiency compared with state-of-the-art supervised learning methods and random selection.

\section{Discussion}
\label{sec:threats}
\subsection{Threats to validity}

As to any empirical study, biases can affect the final results. Therefore, conclusions drawn from this work must be considered with threats to validity in mind. In this section, we discuss the validity of our work.

\textbf{Learner bias.}
This work applies three classifiers, weighted linear SVM, Random Forest and Decision Tree, which are the best setting according to previous research work~\citep{wang2018there}. However, this doesn't necessarily guarantee the best performance in other domains or other static warning datasets. According to the No Free Lunch Theorems~\citep{wolpert1997no}, applying our method framework to other areas would be needed before we can assert that our methods are also better in those domains.

\textbf{Sampling bias.}
One of the most important threats to validity is sampling bias since several sampling methods, random sampling, uncertainty sampling and certainty sampling, are used in combination. However, there are also many sampling methods in the active learning area we can utilize. And different sampling strategies and combinations may result in better performance. This is a potential research direction.

\textbf{Ratio bias.}
In this paper, we propose an ideal scale value for our learner to retrieve on nine static warning datasets to effectively solve the prevalence of false positive in warnings reported by SA tools. obvious improvement is observed for this unbalanced problem. But it doesn't necessarily apply to balanced datasets. 

\textbf{Measurement bias.}
To evaluate the validity of the incremental active learning method proposed in this paper, we employ two measurement metrics: total recall and cost. Several prior research work has demonstrated the necessity and effectiveness of these measurements~\citep{8883076,yu2018finding,yu2019fast2}. Nevertheless, many studies are still based on some classic and traditional metrics, eg. confusion matrix or also known as error matrix~\citep{landgrebe2008efficient}. There exist many popular terminology and derivations from confusion matrix, false positive, F1 score, G measure, and so on. We cannot explore and include all the options in one article. Also, even for this same research methodology, conclusions drawn from different evaluation matrices may differ. However, in this research scenario, it is more efficient to report recall and cost for an effort-aware model.

    
    
    
    

\subsection{Future Work}



\textbf{Estimation.} In real-world problems, labeled data may be scarce or expensive to be obtained, while data without labels may be abundant. In this case, the query process of our incremental learning model cannot safely stop to obtain a given targeted threshold without knowing the actual number of actionable warnings in the data set beforehand. Therefore, estimation is required to guarantee the algorithm stopping detection at an appropriate stage: stopping too late will cause unnecessary cost to explore unactionable warnings and increase false alarms;  while stopping too early may incur missing potential and important true warnings.

\textbf{Ensemble of classifiers.} Ensemble learning is a methodology of making decision based on inputs of multiple experts or classifiers~\citep{zhang2012ensemble}. It's a feasible and important scheme to reduce the variance of classifiers and improve the reliability and robustness of the decision system. The famous No Free Lunch Theorems proposed by Wolpert et al. ~\citep{wolpert1997no} gives us an instinct guidance to recur to ensemble learners. This will be promising to make the best of incremental active learning by precisely making predictions and pinpoint real actionable warnings with a generalized decision system.

\section{Conclusion}
\label{sec:conclusion}

Previous research work shows that about 35\% to 91\% warnings reported as bugs by
static analysis tools are actually unactionable  (i.e., warnings that would not be acted on by developers because they are falsely suggested as bugs). Therefore, to make such systems usable for programmers, some mechanism is required to reduce those false alarms.

Arnold et al.~\citep{arnold2009security} warn that knowledge about what is an ignorable static code warning may not transfer from project to project. Here, they advise that methods for managing static code warnings should be tuned to different software projects.
While we agree with that advice, it does create a knowledge acquisition bottleneck problem since acquiring that knowledge can be a time-consuming and tedious task.
 
This explored methods for acquiring knowledge of what static code warnings can be ignored.
Using a human-in-the-loop active learner,
we conducted an empirical study with 9 software projects and 3 machine learning classifiers to verify how the performance of current SA tools could be improved by an efficient incremental active learning method. We found about 90 \% of actionable static warnings can be identified when only inspecting about 20 \% to 30 \% warning reports without using historical version information.
Our study attempts to bridge the research gap between supervised learning and effort-aware active learning models by an in-depth analysis of reducing the cost of static warning identification problems.

Our method significantly decreases the cost of inspecting falsely reported warnings generated by static code analysis tools for software engineers (especially in the early stage of software project's life cycle) and provides a meaningful guideline to improve the performance of current SA tools. Acceptance and adoption of future static analysis tools can be enhanced by combining with SA feature extraction and self-adaptive incremental active learning.

\section*{Acknowledgements}
This work was partially funded
by NSF grant \#1908762.



\bibliographystyle{model2-names}
\bibliography{main}

\end{document}